\documentclass[graybox]{svmult}

\usepackage{mathptmx}       
\usepackage{helvet}         
\usepackage{courier}        
\usepackage{type1cm}        
\usepackage{makeidx}         
\usepackage{graphicx}        
\usepackage{multicol}        
\usepackage[bottom]{footmisc}

\usepackage{graphicx}
\usepackage{cite}
\usepackage{subfigure}
\usepackage{amsmath}
\usepackage{float}
\usepackage{stfloats}
\usepackage{algorithmic}
\usepackage{algorithm}
\usepackage{amssymb}

\makeindex             

\begin{document}

\title*{On the Spectrum Handoff for Cognitive Radio Ad Hoc Networks without Common Control Channel}
\titlerunning{On the Spectrum Handoff for CR Ad Hoc Networks without Common Control Channel}
\author{Yi Song and Jiang Xie}
\institute{Yi Song \at The University of North Carolina at Charlotte, \email{ysong13@uncc.edu}
\and Jiang Xie \at  The University of North Carolina at Charlotte, \email{Linda.Xie@uncc.edu}}
%
%
\maketitle

\abstract{Cognitive radio (CR) technology is a promising solution to enhance the spectrum utilization by enabling unlicensed users to exploit the spectrum in an opportunistic manner. Since unlicensed users are temporary visitors to the licensed spectrum, they are required to vacate the spectrum when a licensed user reclaims it. Due to the randomness of the appearance of licensed users, disruptions to both licensed and unlicensed communications are often difficult to prevent. In this chapter, a proactive spectrum handoff framework for CR ad hoc networks is proposed to address these concerns. In the proposed framework, channel switching policies and a proactive spectrum handoff protocol are proposed to let unlicensed users vacate a channel \textit{before} a licensed user utilizes it to avoid unwanted interference. Network coordination schemes for unlicensed users are also incorporated into the spectrum handoff protocol design to realize channel rendezvous. Moreover, a distributed channel selection scheme to eliminate collisions among unlicensed users is proposed. In our proposed framework, unlicensed users coordinate with each other without using a common control channel. We compare our proposed proactive spectrum handoff protocol with a reactive spectrum handoff protocol, under which unlicensed users switch channels \textit{after} collisions with licensed transmissions occur. Simulation results show that our proactive spectrum handoff outperforms the reactive spectrum handoff approach in terms of higher throughput and fewer collisions to licensed users. In addition, we propose a novel three dimensional discrete-time Markov chain to characterize the process of reactive spectrum handoffs and analyze the performance of unlicensed users. We validate the numerical results obtained from our proposed Markov model against simulation and investigate other parameters of interest in the spectrum handoff scenario.}

\section{Introduction}
\label{sc:introduction}
The rapid growth of wireless devices has led to a dramatic increase in the need of spectrum access from wireless services. However, according to Federal Communications Commission (FCC) \cite{FCC-2003}, up to 85\% of the assigned spectrum is underutilized due to the current fixed spectrum allocation policy. In order to overcome the imbalance between the increase in the spectrum access demand and the inefficiency in the spectrum usage, FCC has suggested a new paradigm for dynamically accessing the assigned spectrum where the spectrum is not used \cite{FCC-docket2003}. Cognitive radio (CR) is a key technology to realize dynamic spectrum access (DSA) that enables an unlicensed user (or, secondary user) to adaptively adjust its operating parameters and exploit the spectrum which is unused by licensed users (or, primary users) in an opportunistic manner \cite{Mitola00}.

The CR technology allows secondary users (SUs) to seek and utilize ``spectrum holes'' in a time and location-varying radio environment without causing harmful interference to primary users (PUs). This opportunistic use of the spectrum leads to new challenges to make the network protocols adaptive to the varying available spectrum \cite{Akyildiz-Lee06}. Specifically, one of the most important functionalities of CR networks is \textit{spectrum mobility}, which enables SUs to change the operating frequencies based on the availability of the spectrum. Spectrum mobility gives rise to a new type of handoff called \textit{spectrum handoff}, which refers to the process that when the current channel used by a SU is no longer available, the SU needs to pause its on-going transmission, vacate that channel, and determine a new available channel to continue the transmission. Compared with other functionalities (\textit{spectrum sensing}, \textit{spectrum management}, and \textit{spectrum sharing}) \cite{Akyildiz-Lee06} of CR networks, spectrum mobility is less explored in the research community. However, due to the randomness of the appearance of PUs, it is extremely difficult to achieve fast and smooth spectrum transition leading to minimum interference to legacy users and performance degradation of secondary users during a spectrum handoff. This problem becomes even more challenging in ad hoc networks where there is no centralized entity (e.g., a spectrum broker \cite{Akyildiz-Lee06}) to control the spectrum mobility.

\subsection{Spectrum Handoff in Cognitive Radio Networks}
\label{ssc:handoff}
Related work on spectrum handoffs in CR networks falls into two categories based on the moment when SUs carry out spectrum handoffs. One approach is that SUs perform spectrum switching and radio frequency (RF) front-end reconfiguration \textit{after} detecting a PU \cite{Willkomm05,Mishra-ICC-06,LCWang09,Zhang09,CWWangGC10}, namely the \textit{reactive} approach. Although the concept of this approach is intuitive, there is a non-negligible sensing and reconfiguration delay which causes unavoidable disruptions to both the PU and SU transmissions. Another approach is that SUs predict the future channel availability status and perform spectrum switching and RF reconfiguration \textit{before} a PU occupies the channel based on observed channel usage statistics\cite{Zheng-proactive08,Clancy-2006,Arslan-predict07,Yoon10icc}, namely the \textit{proactive} approach. This approach can dramatically reduce the collisions between SUs and PUs by letting SUs vacate channels before a PU reclaims the channel. In the existing proposals of the proactive approach, a predictive model for dynamic spectrum access based on the past channel usage history is proposed in \cite{Zheng-proactive08}. A cyclostationary detection and Hidden Markov Models for predicting the channel idle times are proposed in \cite{Clancy-2006}. In \cite{Arslan-predict07}, a binary time series for the spectrum occupancy characterization and prediction is proposed. In \cite{Yoon10icc}, a novel spectrum handoff scheme called voluntary spectrum handoff is proposed to minimize SU disruption periods during spectrum handoffs. In \cite{CSongINFOCOM10}, the error of prediction of the channel usage is considered in designing an intelligent dynamic spectrum access mechanism. In \cite{GeirhoferMobicom09}, an experimental cognitive radio test bed is presented. It uses sensing and channel usage prediction to exploit temporal white space between primary WLAN transmissions.

\subsection{Common Control Channel in Cognitive Radio Networks}
\label{ssc:ccc}
A common control channel (CCC) is used for supporting the network coordination and channel related information exchange among SUs. In the prior proposals of the above two spectrum handoff approaches, the network coordination and rendezvous issue (i.e., before transmitting a packet between two nodes, they first find a common channel and establish a link) is either not considered\cite{LCWang09}\cite{Zhang09}\cite{Clancy-2006}\cite{Arslan-predict07}\cite{CSongINFOCOM10}\cite{GeirhoferMobicom09} or simplified by using a global common control channel (CCC)\cite{Willkomm05}\cite{Mishra-ICC-06}\cite{Zheng-proactive08}\cite{Yoon10icc}. A SU utilizing a channel without coordinating with other SUs may lead to the failure of link establishment \cite{Akyildiz09-LC}. Therefore, network coordination has significant impact on the performance of SUs. Although a global CCC simplifies the network coordination among SUs \cite{Kond08}, there are several limitations when using this approach in CR networks. First of all, it is difficult to identify a global CCC for all the secondary users throughout the network since the spectrum availability varies with time and location. Secondly, the CCC is influenced by the primary user traffic because a PU may suddenly appear on the current control channel. For these reasons, IEEE 802.22 \cite{Stevenson09}, the first standard based on the use of cognitive radio technology on the TV band between 41 and 910 MHz, does not utilize a dedicated channel for control signaling, instead dynamically choosing a channel which is not used by legacy users \cite{Chal06wicon}.  

In this chapter, we investigate the network scenario where no CCC exists and its impact on the spectrum handoff design in CR ad hoc networks. Since when no CCC exists in the network, message exchange among SUs is not always feasible. Thus, the spectrum handoff design becomes more challenging than the scenario with a CCC. Currently, several proposals have been proposed to accomplish network coordination without a CCC in ad hoc networks. Based on the number of users making link agreements simultaneously, the proposed network coordination schemes can be categorized into (1) single rendezvous coordination schemes \cite{Mo-MAC2008,Tzamaloukas00-1,Tzamaloukas00-2} (i.e., only one pair of SUs in a network can exchange control information and establish a link at one time) and (2) multiple rendezvous coordination schemes \cite{So07mcmac,Bahl04ssch:slotted,Theis10} (i.e., multiple pairs of SUs in a network can use different channels to exchange control information and establish multiple links at the same time). Thus, we utilize these two types of network coordination schemes and incorporate them into the spectrum handoff design for CR ad hoc networks.

\subsection{Channel Selection in Cognitive Radio Networks}
\label{ssc:channel selection}
Even though the channel allocation issue has been well studied in traditional wireless networks (e.g., cellular networks and wireless local area networks (WLANs)), channel allocation in CR networks, especially in a spectrum handoff scenario, still lacks sufficient research. When SUs perform spectrum handoffs, a well-designed channel selection method is required to provide fairness for all SUs as well as to avoid multiple SUs to select the same channel at the same time. Currently, the channel selection issue in a multi-user CR network is investigated mainly using game theoretic approaches \cite{Neel-phdthesis,J_Sen08,J_Niy08,Zheng05,Zhang02}, while properties of interest during spectrum handoffs, such as SU handoff delay and SU service time, are not studied. Furthermore, most of the prior work on channel allocation in spectrum handoffs \cite{LCWang09}\cite{Zheng-proactive08} only considers a two-secondary-user scenario, where a SU greedily selects the channel which either results in the minimum service time \cite{LCWang09} or has the highest probability of being idle \cite{Zheng-proactive08}. In \cite{Yoon10icc}, only one pair of SUs is considered and the channel selection issue is ignored. However, if multiple SUs perform spectrum handoffs at the same time, these channel selection methods will cause definite collisions among SUs. Hence, the channel selection method aiming to prevent collisions among SUs in a multi-secondary-user spectrum handoff scenario is ignored in the prior work.

\subsection{Analytical Model for Spectrum Handoff in Cognitive Radio Networks}
\label{ssc:analytical}
An analytical model is of great importance for performance analysis because it can provide useful insights on the operation of spectrum handoffs. However, there have been limited studies on the performance analysis of spectrum handoffs in CR networks using analytical models. The performance analysis of all prior works on spectrum handoffs is simulation-based with the exception of \cite{LCWang09} and \cite{CWWangGC10}. In \cite{LCWang09} and \cite{CWWangGC10}, a preemptive resume priority queueing model is proposed to analyze the total service time of SU communications for proactive and reactive-decision spectrum handoffs. However, in both \cite{LCWang09} and \cite{CWWangGC10}, only one pair of SUs is considered in a network, while the interference and interactions among SUs are ignored, which may greatly affect the performance of the network. In all the above proposals, a common and severe limitation is that the authors assume that the detection of PUs is perfect (i.e., a SU transmitting pair can immediately perform channel switching if a PU is detected to appear on the current channel, thus the overlapping of SU and PU transmissions is negligible). However, since the power of a transmitted signal is much higher than the power of the received signal in wireless medium due to path loss, instantaneous collision detection is not possible for wireless communications \cite{Tang07}. Thus, even if only a portion of a packet is collided with another transmission, the whole packet is wasted and need to be retransmitted. Without considering the retransmission, the performance conclusion may be inaccurate, especially in wireless communications. Unfortunately, it is not easy to simply add retransmissions in the existing models. In this chapter, we model the retransmissions of the collided packets in our proposed Markov model. 

\subsection{Contributions}
\label{ssc:contribution}
This chapter studies the spectrum handoff issues in cognitive radio networks without the existence of a CCC. The contributions of our work are as follows:
\renewcommand{\labelitemi}{$\blacksquare$}
\begin{itemize} \item Due to the spectrum-varying nature of CR networks, we consider more practical coordination schemes instead of using a CCC to realize channel rendezvous. We incorporate two types of channel rendezvous and coordination schemes into the spectrum handoff design and compare the performance of our proposed spectrum handoff protocol with the reactive spectrum handoff approach under different coordination schemes. 
\item Based on the observed channel usage statistics, we propose proactive spectrum handoff criteria and policies for SUs using a probability-based prediction method. SUs equipped with the prediction capability can proactively predict the idleness probability of the spectrum band in the near future. Thus, harmful interference between SUs and PUs can be diminished and SU throughput is increased. In addition, by considering channel rendezvous and coordination schemes, we propose a proactive spectrum handoff protocol for SUs based on our proposed handoff criteria and policies. 
\item With the aim of eliminating collisions among SUs and achieving short spectrum handoff delay, we propose a novel distributed channel selection scheme especially designed for multi-user spectrum handoff scenarios. Our proposed channel selection scheme does not involve centralized controller and only need SUs to broadcast their sensed channel availability information once, which drastically reduces the message exchange overhead. 
\item We propose a novel three dimensional discrete-time Markov model to characterize the process of reactive spectrum handoffs and analyze the performance of SUs. We implement one of the considered network coordination schemes in our model. Since instantaneous collision detection is not feasible for wireless communications, we consider the retransmissions of the collided SU packets in spectrum handoff scenarios. We also consider the spectrum sensing delay and its impact on the network performance.\end{itemize}

\subsection{Organization}
\label{ssc:organization}
The rest of this chapter is organized as follows. In Section \ref{sc:networkmodel}, network coordination schemes and assumptions considered in this chapter are introduced. In Section \ref{sc:protocol}, the details of the proposed proactive spectrum handoff framework are given. In Section \ref{sc:selection}, the algorithm of the proposed distributed channel selection scheme is presented. Simulation results of our proposed spectrum handoff framework are presented in Section \ref{sc:evaluation}. In Section \ref{sc:analysis}, a three dimensional discrete-time Markov model is proposed, followed by the conclusions in Section \ref{sc:conclusion}.

\section{Network Coordination and Assumptions}
\label{sc:networkmodel}
\subsection{Single Rendezvous Coordination Scheme}
\label{ssc:singlecoordination}
\begin{figure}[htb]
\centerline{\includegraphics[width=0.85\textwidth]{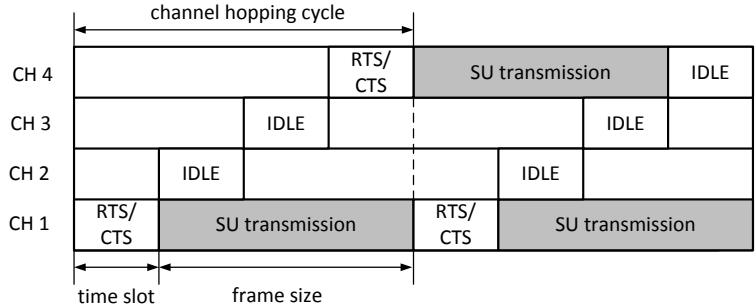}}
\caption{An example of the single rendezvous coordination scheme.}
\label{fig:coordination}
\end{figure}
We consider a network scenario where $N$ SUs form a CR ad hoc network and opportunistically access $M$ orthogonal licensed channels. For the single rendezvous coordination scheme, we use Common Hopping as the channel coordination scheme \cite{Mo-MAC2008}. Fig. \ref{fig:coordination} illustrates the operations of Common Hopping, under which the channels are time-slotted and SUs communicate with each other in a synchronous manner. This is similar to the frequency hopping technique used in Bluetooth \cite{Ferro05}. When no packet needs to be transmitted, all the SU devices hop through channels using the same hopping sequence (e.g., the hopping pattern cycles through channels $1,2,\cdots,M$). The length of a time slot (i.e., the dwelling time on each channel during hopping) is denoted as $\beta$. If a pair of SUs wants to initiate a transmission, they first exchange request-to-send (RTS) and clear-to-send (CTS) packets during a time slot. Then, after the SU transmitter successfully receives the CTS packet, they pause the channel hopping and remain on the same channel for data transmissions, while other non-transmitting SUs continue hopping. After the data being successfully transmitted, the SU pair rejoins the channel hopping. 
\subsection{Multiple Rendezvous Coordination Scheme}
\label{ssc:multiplecoordination}
\begin{figure}[htb]
\centerline{\includegraphics[width=0.83\textwidth]{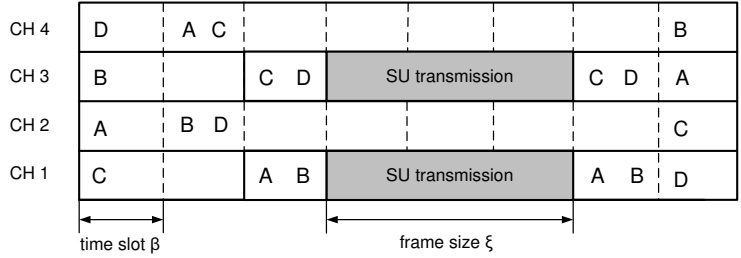}}
\caption{An example of the multiple rendezvous coordination scheme.}
\label{fig:multicoordination}
\end{figure}
Unlike in the single rendezvous coordination scheme that only one pair of SUs can make an agreement in one time slot, in the multiple rendezvous coordination scheme, multiple SU pairs can make agreements simultaneously on different channels. A typical example of this type of coordination schemes is McMAC \cite{So07mcmac}. Fig. \ref{fig:multicoordination} depicts the operations of McMAC. Instead of using the same channel hopping sequence for all the SUs, in McMAC, each SU generates a distinct pseudo-random hopping sequence (e.g., in Fig. \ref{fig:multicoordination}, the channel hopping sequence for user \textit{A} is 2-4-1-3, for user \textit{B} is 3-2-1-4, etc.). When a SU is idle, it follows its default hopping sequence to hop through the channels. If a SU intends to send data to a receiver, it temporarily tunes to the current channel of the receiver and sends a RTS during the time slot (i.e., in Fig. \ref{fig:multicoordination}, SUs $A$$B$ and $C$$D$ are two transmitting pairs that intend to initiate new transmissions at the same time). Then, if the receiver replies with a CTS, both the transmitter and the receiver stop channel hopping and start a data transmission on the same channel. When they finish the data transmission, they resume to their default channel hopping sequences. In this chapter, we consider the scenario where SU nodes are aware of each other's channel hopping sequences \cite{So07mcmac}.

In this chapter, we assume that stringent time synchronization among SUs for channel hopping can be achieved without the need to exchange control messages on a CCC in both cases. We consider a synchronization scheme similar to the one used in \cite{So07mcmac} that every SU includes a time stamp in every packet it sends. Then, a SU transmitter obtains the clock information of the intended SU receiver by listening to the corresponding channel and estimates the rate of clock drift to realize time synchronization. Various schemes have been proposed to calculate the rate of clock drift for synchronization \cite{Siv04}. 

In both types of coordination schemes, we assume that any SU data packet is transmitted at the beginning of a time slot and ends at the end of a time slot. This implies that the length of a SU data packet, $\delta$, is a multiple of the time slot. This assumption is commonly used in time-slotted systems \cite{Su08}\cite{Su-ciss-08}. We further define that a SU data packet is segmented into frames and each frame contains $c$ time slots. The length of a frame is denoted as $\xi$, so $\xi=c\beta$. As shown in Fig. \ref{fig:coordination}, at the end of a frame, the two SUs can either rejoin the channel hopping when a data transmission ends, or start another data transmission by exchanging RTS/CTS packets.

\subsection{Network Assumptions}
\label{ssc:channelmodel}
\begin{figure}[htb!]
\centerline{\includegraphics[width=0.75\textwidth]{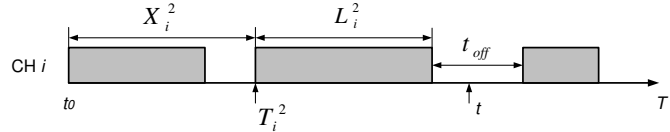}}
\caption{The PU traffic activity on channel $i$.}
\label{fig:prediction}
\end{figure}

In this chapter, we model each licensed channel as an ON-OFF process \cite{Tang08}\cite{Arslan-predict07}. As shown in Fig. \ref{fig:prediction}, each rectangle represents a PU data packet being transmitted on a channel (i.e., the ON period) and the other blank areas represent the idle periods (i.e., the OFF period). The length of a rectangle indicates the packet length of a PU data packet. Therefore, a SU can only utilize a channel when no PU transmits at the same time. In Fig. \ref{fig:prediction}, $t_0$ represents the time a SU starts channel prediction. Thus, for the $i$-th channel at any future time $t~ (t>t_0)$, the status of the channel is denoted as $N_i(t)$ which is a binary random variable with values 0 and 1 representing the idle and the busy state, respectively. We also assume that each PU is an $M/G/1$ system \cite{LCWang09}\cite{Wang-Wang08}, that is, the PU packet arrival process follows the Poisson process with the average arrival rate $\bar{\lambda}_i$ and the length of a data packet follows an arbitrary probability density function (pdf) $f_{L_i}(l)$.

Due to the fact that the power of a transmitted signal is much higher than the power of the received signal in wireless medium, instantaneous collision detection is not possible for wireless nodes. Thus, we assume that if a SU frame collides with a PU packet, the wasted frame can only be retransmitted at the end of the frame. In addition, in our proposed spectrum handoff protocol, we assume that each SU is equipped with two radios. One is used for data and control message transmission, namely the transmitting radio. The other is applied to scan all the channels in the band and to obtain the channel occupancy information, namely the scanning radio. The scanning radio has two major functions for the proposed protocol: (1) observe the channel usage and store the channel statistics in the memory for future channel availability prediction and (2) confirm that the newly selected channel is idle for SU transmissions.
\vspace{-0.2in}

\section{Proactive Spectrum Handoff Protocol}
\label{sc:protocol}
\subsection{Proposed Spectrum Handoff Criteria and Policies}
\label{ssc:swiching}
By utilizing the observed channel usage statistics, a SU can make predictions of the channel availability before the current transmission frame ends. Based on the prediction, the SU decides whether to stay in the present channel, or switch to a new channel, or stop the on-going transmission. We propose two criteria for determining whether a spectrum handoff should occur: (1) the predicted probability that the current and a candidate channel (i.e., a channel that can be selected for continuing the current data transmission) is busy or idle and (2) the expected length of the channel idle period. Based on these criteria, we design spectrum handoff policies.

Fig. \ref{fig:prediction} shows the PU user traffic activity on channel $i$, where $X_i^k$ and $T_i^k$ represent the inter-arrival time and arrival time of the $k$-th packet, respectively. Consistent with the assumption that PU packets arrive in a Poisson stream fashion \cite{Wang-Wang08}, $X_i^k$ is exponentially distributed with the average arrival rate $\bar{\lambda}_i$ packets per second and the PU packet length follows the pdf $f_{L_i}(l)$. According to Fig. \ref{fig:prediction}, for any future time $t$, the probability that the $i$-th channel is busy or idle can be written as follows:
\begin{equation}\label{eq:probability}
\begin{array}{llr}
\Pr(N_i(t)=1) \text{ if~~} &T_i^k<t \text{~and } T_i^k+L_i^k \geq t, ~~~~~~k \geq 1, \\
\Pr(N_i(t)=0) \text{ if~~} &T_i^k+L_i^k <t \text{ and } T_i^{k+1} \geq t, ~~k \geq 1\\
& T_i^{k+1} \geq t, ~~~~~~~~~~~~~~~~~~~~~~~~~~~~k = 0,
\end{array}
\end{equation}
where $L_i^k$ denotes the length of the $k$th PU data packet on channel $i$. Therefore, the probability that channel $i$ is idle at any future time $t$ can be obtained by (\ref{eq:channelidle}).
\begin{equation}
\begin{split}
\Pr(N_i(t)\!\!\!=\!\!0)&\!\!=\!\!\int_{0}^{\infty}\!\!\left[\!\sum_{k\!=\!1}^\infty \!\Pr(\!T_i^k\!\!+\!L_i\!<\!t|k)\Pr(\!T_i^{k\!+\!1}\!\geq\!t|k)\Pr(k)\!\!+\!\Pr(\!T_i^1\!\geq\! t)\Pr(k\!\!=\!0)\!\right]\!\!f_{L_i}(l)\,\mathrm{d}l \\
  \!\!=&\!\!\int_{0}^{\infty}\!\!\left\lbrace \sum_{k\!=\!1}^\infty \!\left[\!\frac{\bigl(\bar{\lambda}_i(t\!-\!L_i)\bigr)^{k}}{k!}\!e^{-\bar{\lambda}_i(t\!-\!L_i)}\!\right]\! \left(\!\frac{(\bar{\lambda}_it)^k}{k!}\!e^{-\bar{\lambda}_it}\!\right)\!\frac{(\bar{\lambda}_it)^k}{k!}\!e^{-\bar{\lambda}_it}\!+\!e^{-2\bar{\lambda}_it}\!\right\rbrace\! f_{L_i}(l)\,\mathrm{d}l.
\end{split}
\label{eq:channelidle}
\end{equation}
Let $t_{off}$ represent the duration of the OFF period. For the $i$-th channel, the cumulative distribution function (CDF) of the duration of the OFF period is:
\begin{equation}\label{eq:offperiod}
\begin{split}
\Pr(t_{off}<x) & = \int_{0}^{\infty} \int_{0}^{l+x} \bar{\lambda}_ie^{-\bar{\lambda}_i t}f_{L_i}(l)\, \mathrm{d}t\mathrm{d}l \\
& =\int_{0}^{\infty}\left(1-e^{-\bar{\lambda}_i(l+x)}\right)f_{L_i}(l)\, \mathrm{d}l.
\end{split}
\end{equation}

Hence, based on the above prediction, the policy that a SU should switch to a new channel is:
\begin{equation}\label{eq:switch}
\Pr(N_i(t)=0)<\tau_L,
\end{equation}
where $\tau_L$ is the probability threshold below which a channel is considered to be busy and the SU needs to carry out a spectrum handoff, that is, the current channel is no longer considered to be idle at the end of the frame transmission. In addition, the policies that a channel $j$ becomes a candidate channel at time $t$ are:
\begin{equation}\label{eq:candidate}
\left\{
\begin{array}{l}
\Pr(N_j(t)=0) \geq \tau_H \\
\Pr(t_{j,off}>\eta) \geq \theta,
\end{array}
\right.
\end{equation}
where $\tau_H$ is the probability threshold for a channel to be considered idle at the end of the current frame, $\eta$ is the length of a frame plus a time slot (i.e., $\eta=\xi+\beta$), and $\theta$ is the probability threshold for a channel to be considered idle for the next frame transmission. The second criterion in (\ref{eq:candidate}) means that, in order to support at least one SU frame, the probability that the duration of the idleness of the $j$-th channel to be longer than a frame size must be higher than or equal to $\theta$. 

\subsection{Proposed Spectrum Handoff Protocol Details}
\label{ssc:protocol}
The proposed spectrum handoff protocol is based on the above proposed spectrum handoff policies. It consists of two parts. The first part, namely \textit{Protocol 1} (the pseudo code of Protocol 1 is presented in Algorithm \ref{al:protocol1}\footnote{DAT is the flag for data transmission requests, DSF is the data-sending flag, $t$ is the beginning of the next slot, and $k$ is the next hopping channel in the single rendezvous coordination scheme or the hopping channel for the receiver in the multiple rendezvous coordination scheme.}), describes how a SU pair initiates a new transmission. Regardless of the coordination schemes used during channel hopping, if a data packet arrives at a SU, the SU predicts the availability of the next hopping channel (in the single rendezvous coordination scheme case) or the hopping channel of the receiver (in the multiple rendezvous coordination scheme case) at the beginning of the next slot. Based on the prediction results, if the channel satisfies the policies in (\ref{eq:candidate}) for data transmissions, the transmitter sends a RTS packet to the receiver on the same hopping channel as the receiver at the beginning of the next time slot. Upon receiving the RTS packet, the intended SU receiver replies a CTS packet in the same time slot. Then, if the CTS packet is successfully received by the SU transmitter, the two SUs pause the channel hopping and start the data transmission on the same channel. Note that if more than one pair of SUs contend the same hopping channel for data transmission, an algorithm that eliminates SU collisions is proposed in Section \ref{sc:selection}.
\begin{algorithm}
\caption{Protocol 1: starting a new transmission}
\label{al:protocol1}
\noindent \textbf{Register initiation}: DAT:=0, DSF:=0; \\
predicting $\Pr(N_k(t)=0)$, $\Pr(t_{k,off}>\eta)$;\\
\textbf{if} $\Pr(N_k(t)=0) \geq \tau_H$ AND $\Pr(t_{k,off}>\eta) \geq \theta$ \\
\indent \hspace{0.1in} DAT := 1; \\
\textbf{end if} \\
\textbf{if} DAT=1\\
\indent \hspace{0.1in} sending RTS; \\
\textbf{end if} \\
\textbf{upon} receiving CTS \\
\indent \hspace{0.1in} DSF := 1; \\
\textbf{if} DSF=1 \\
\indent \hspace{0.1in} DSF := 0; \\
\indent \hspace{0.1in} transmitting a data frame;\\
\indent \hspace{0.1in} DAT := 0 when transmission ends; \\
\textbf{end if}
\end{algorithm}

The second part, namely \textit{Protocol 2} (the pseudo code of Protocol 2 is presented in Algorithm \ref{al:protocol2}\footnote{CSW is the channel switching flag, NUC and LSC are the number and the list of the candidate channels for data transmissions, respectively, and channel $i$ is the current channel. As similar in Protocol 1, DAT is the flag for data transmission requests and DSF is the data-sending flag.}), is on the proactive spectrum handoff during a SU transmission. The goal of our proposed protocol is to determine whether the SU transmitting pair needs to carry out a spectrum handoff and then switch to a new channel by the time a frame transmission ends. Using the proposed protocol, the SU transmitting pair can avoid disruptions with PUs when PUs appear.
\begin{algorithm}
\caption{Protocol 2: spectrum handoff during a transmission}
\label{al:protocol2}
\noindent \textbf{Register initiation}: CSW:=0, DSF:=0, NUC:=0, LSC:=$\emptyset$; \\
\textbf{for} $j:=0, j\leq M$ \textbf{do} \\
\indent \hspace{0.1in} predicting $\Pr(N_j(t)=0)$, $\Pr(t_{j,off}>\eta$);\\
\textbf{end for} \\
\textbf{if} $\Pr(N_i(t)=0)<\tau_L$ AND DAT=1 \\
\indent \hspace{0.1in} CSW := 1; \\\textbf{end if}\\
\textbf{if} CSW=1 \\
\indent \hspace{0.1in} \textbf{for} $k:=0, k\leq M$ \textbf{do} \\
\indent \hspace{0.3in} \textbf{if} $\Pr(N_k(t)=0) \geq \tau_H$ AND $\Pr(t_{off}>\eta) \geq \theta$ \\
\indent \hspace{0.5in} NUC := NUC+1; \\
\indent \hspace{0.5in} LSC(NUC) := $k$; \\
\indent \hspace{0.3in} \textbf{end if} \\
\indent \hspace{0.1in} \textbf{end for} \\
\textbf{end if}\\
\textbf{if} LSC=$\emptyset$\\
\indent \hspace{0.1in} transmission stops and launch Protocol 2;\\
\textbf{elseif} LSC $\neq \emptyset$ \\
\indent \hspace{0.1in} start scanning radio; \\
\indent \hspace{0.1in} launch channel selection algorithm in LSC; \\
\indent \hspace{0.1in} sending CSR; \\
\textbf{end if} \\
\textbf{upon} receiving CSA \textbf{then}\\
\indent \hspace{0.1in} switch to the selected channel and start scanning radio; \\
\textbf{if} channel is busy \\
\indent \hspace{0.1in} transmission stops and launch Protocol 2;\\
\textbf{else} DSF := 1 CSW:=0; \\
\textbf{end if} \\
\textbf{if} DSF=1 \\
\indent \hspace{0.1in} DSF := 0; \\
\indent \hspace{0.1in} transmitting a data frame; \\
\indent \hspace{0.1in} DAT := 0 when transmission ends; \\
\textbf{end if}
\end{algorithm}

Based on the observed channel usage information, a SU transmitter checks the spectrum handoff policy in (\ref{eq:switch}) for the current channel by predicting the channel availability at the end of the frame. If the policy is not satisfied, this means that the current channel is still available for the next frame transmission. Then, the SU transmitting pair does not perform a spectrum handoff and keeps staying on the same channel. However, if the policy is satisfied, the \textit{channel-switching} (CSW) flag is set, that is, the current channel is considered to be busy during the next frame time and the SUs need to perform a spectrum handoff by the end of the frame to avoid harmful interference to a PU who may use the current channel. After the CSW is set, the two SUs rejoin the channel hopping in the next time slot after the previous frame. In the proposed distributed channel selection algorithm (which is explained in detail in Section \ref{sc:selection}), the SUs that need to perform spectrum handoffs at the same time are required to update the predicted channel availability information to other SUs. Hence, the SUs need to hop to the same channel to inform neighboring SUs. Note that in the single rendezvous coordination scheme, all SUs that do not transmit data follow the same hopping sequence. Therefore, when the CSW flag is set, all SUs that need to perform a spectrum handoff pause the current transmission and resume the channel hopping with the same sequence, so they will hop to the same channel. However, in the multiple rendezvous coordination scheme, each SU follows a default hopping sequence which may not be the same as other's hopping sequence. In order to be able to exchange channel availability information among SUs on the same channel, in our proposed protocol, SUs are required to follow the same hopping sequence only when performing spectrum handoffs.

On the other hand, the SU transmitter checks the criteria in (\ref{eq:candidate}) for available handoff candidate channels in the band. If no channel is available, then the on-going transmission stops immediately at the end of the frame. The two SUs hop to the next channel for one more time slot and check the channel availability based on the criteria in (\ref{eq:candidate}) at the beginning of the next time slot for both the single rendezvous and the multiple rendezvous coordination schemes. However, if the set of the handoff candidate channels is not empty, the SU transmitter triggers a distributed channel selection algorithm (which is explained in detail in Section IV) and sends a \textit{channel-switching-request} (CSR) packet containing the newly selected channel information in the next time slot. Upon receiving the CSR packet, the SU receiver replies with a \textit{channel-switching-acknowledgement} (CSA) packet. If the CSA packet is successfully received by the SU transmitter, this means that the channel switching agreement between the two SU nodes has been established. Thus, both SU nodes switch to the selected channel and start the data transmission for the next frame. The handoff delay of a spectrum handoff is defined as the duration from the time a SU vacates the current channel to the time it resumes the transmission. Note that there is a possibility that the prediction is not correct and there is a PU on the channel which the SUs switch to. Hence, at the beginning of the frame, the SU transmitting pair restarts the scanning radio to confirm that the selected channel is idle. If the channel is sensed busy, the two SUs immediately resume the channel hopping and launch Protocol 2.

\section{Distributed Channel Selection Algorithm}
\label{sc:selection}
\subsection{Procedure of the Proposed Channel Selection Algorithm}
\label{scc:procedure}
The channel selection issue should be handled with caution to avoid collisions among SUs. On one hand, preventing SU collisions is more important in the spectrum handoff scenario than in general channel allocation scenarios \cite{Zheng05} due to the fact that collisions among SUs lead to data transmission failures, thus they may result in long spectrum handoff delay, which has deteriorating effect on delay-sensitive network applications. Additionally, the channel selection algorithm also should be executed fast in order to achieve short handoff delay. On the other hand, since no centralized network entity exists in CR ad hoc networks to manage the spectrum allocation, the channel selection algorithm should be applied in a distributed manner to prevent SU collisions.

Our goal is to design a channel selection scheme for the spectrum handoff scenario in CR ad hoc networks that can eliminate collisions among SUs in a distributed fashion. Based on the protocols described in Section \ref{ssc:protocol}, there are two cases in preventing collisions among SUs. The first case is that during the channel hopping phase, if more than one SU transmitters want to initiate new data transmissions, a collision occurs when they send RTS packets on the same channel at the same time, namely the type 1 collision. The second case is that when more than one SU pairs perform spectrum handoffs at the same time, a collision occurs when they select the same channel to switch to, namely the type 2 collision. Once a collision happens, all packets involved are wasted and need to be retransmitted. Since the spectrum handoff delay of an on-going transmission is more critical than the packet waiting time of a new transmission (i.e., the duration from the time a new packet arrives until it is successfully transmitted), the type 2 collision should be prevented with higher priority than the type 1 collision. 
\begin{figure}[hbt]
\centerline{\includegraphics[width=0.6\textwidth]{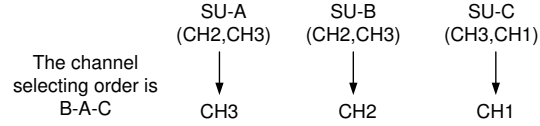}}
\caption{An example of the proposed channel selection scheme.}
\label{fig:selection}
\end{figure}

Fig. \ref{fig:selection} describes an example of the proposed channel selection scheme, where three SUs, \textit{A}, \textit{B}, and \textit{C}, perform spectrum handoffs at the same time. In the parenthesis, the candidate channels are ordered based on the criterion for channel selection (e.g., the probability that a channel is idle). The proposed channel selection procedure is summarized as follows:

\textit{Step 1} \textit{Pseudo-random Sequence Generation}: At each time slot, a pseudo-random channel selecting sequence is generated locally that all SU transmitters involved in spectrum handoffs should follow to choose channels. In Fig. \ref{fig:selection}, the channel selecting sequence for all SUs is \textit{B-A-C}. Since the sequence is generated with the same seed (e.g., the time stamp), every SU generates the same channel selecting sequence at the same time slot. However, the selecting sequences are different at different time slots.

\textit{Step 2} \textit{Channel Information Update}: For both the single rendezvous coordination scheme and the multiple rendezvous coordination scheme, all SUs follow the same sequence to hop through the channels during spectrum handoffs. Hence, when a SU needs to perform a spectrum handoff at the beginning of a time slot, it broadcasts the sensed channel availability information to neighboring SU nodes on the current hopping channel if it is idle. To avoid collisions of the broadcast messages, a time slot is further divided into $W$ mini slots, $W$ is an integer defined by the system. A SU broadcasts the channel availability information only in the corresponding mini slot based on the selecting sequence generated in Step 1. In the example shown in Fig. \ref{fig:selection}, SU-B broadcasts the channel availability information in the first mini slot, SU-A broadcasts in the second mini slot, and SU-C broadcasts in the third mini slot. If the broadcasting process cannot finish within one time slot due to many SUs performing spectrum handoffs at the same time, it should continue in the next time slot until all SUs broadcast the channel information messages. Hence, a SU can obtain the channel availability information predicted by all the neighboring SUs who need to perform spectrum handoffs.

\textit{Step 3} \textit{Channel Selection}: Every SU who needs to perform a spectrum handoff computes the target handoff channel for its spectrum handoff based on the selecting sequence and the criterion for channel selection. The pseudo code of the algorithm for computing the target channel is presented in Algorithm \ref{al:targetchannel}, where $C_i$ denotes the target handoff channel for $SU_i$. In the example shown in Fig. \ref{fig:selection}, based on the selecting sequence, SU-B selects the first channel (i.e., channel 2) in its available channel list. Thus, the remaining SUs delete channel 2 in their available channel lists. Then, SU-A selects channel 3 so on and so forth. Therefore, for each SU, the proposed channel selection algorithm terminates until an available channel is selected or all available channels are depleted. If the target channel exists, then the SU selects it to resume its data transmission; otherwise, the SU waits for the next time slot to perform the spectrum handoff. Since the selecting sequence and the channel availability information are known to every SU who perform the spectrum handoff at the same time, the target channel for each SU (i.e., $C_k, k\in[1,N]$) is also known. Thus, the collision among SUs can be avoided.
\vspace{-0.1in}
\begin{algorithm}
\caption{Computing the Target Channel for SU $k$}
\label{al:targetchannel}
\textbf{Input:} selecting sequence $s$, the list of candidate channels $l_n,n\in[1,N]$\\
\textbf{Output:} target channel $C_k$\\
\textbf{for} $i:=1, i\leq N$ \textbf{do}     \hspace{0.4in}// starting from the first SU in $s$\\
\indent \hspace{0.1in}\textbf{if} $s(i) \ne k$  \\
\indent \hspace{0.2in}\textbf{if} $l_{s(i)} = \emptyset$ \hspace{0.65in}// if the list of candidate channels of SU $s(i)$ is empty\\
\indent \hspace{0.3in}$C_{s(i)} := NULL$ \\
\indent \hspace{0.2in}\textbf{elseif} $l_{s(i)} \neq \emptyset$ \hspace{0.45in}// if the list of candidate channels of SU $s(i)$ is not empty\\
\indent \hspace{0.3in}$C_{s(i)} := \arg\max_{j \in l_{s(i)}}(\Pr(N_j(t)=0))$\\
\indent \hspace{0.2in}\textbf{end if} \\
\indent \hspace{0.2in}\textbf{for} $m:=i+1, m\leq N$ \textbf{do} \\
\indent \hspace{0.3in}\textbf{if} $C_{s(i)} \in l_{s(m)}$ \hspace{0.4in}// if $C_{s(i)}$ is in the list of candidate channels of SU $s(m), i\!<\!m\!\leq\!N$\\
\indent \hspace{0.4in}$l_{s(m)} := l_{s(m)} - C_{s(i)}$ \hspace{0.1in}// remove the channel from the list\\
\indent \hspace{0.3in}\textbf{end if} \\
\indent \hspace{0.2in}\textbf{end for} \\
\indent \hspace{0.1in}\textbf{elseif} $s(i) = k$ \\
\indent \hspace{0.2in}\textbf{if} $l_k = \emptyset$ \hspace{0.65in}// if the list of candidate channels of SU $k$ is empty\\
\indent \hspace{0.3in}\textbf{return} $C_k := NULL$ \textbf{break} \hspace{0.3in}// no available channel for SU $k$ \\
\indent \hspace{0.2in}\textbf{elseif} $l_k \neq \emptyset$ \hspace{1.1in}// if the list of candidate channels of SU $k$ is not empty\\
\indent \hspace{0.3in}\textbf{return} $C_k := \arg\max_{j \in l_k}(\Pr(N_j(t)=0))$ \textbf{break}\\
\indent \hspace{1.2in} // SU $k$ selects the channel that has the highest probability of being idle\\
\indent \hspace{0.2in}\textbf{end if} \\
\indent \hspace{0.1in}\textbf{end if} \\
\textbf{end for}
\end{algorithm}
\vspace{-0.2in}

\subsection{Fairness and Scalability of the Proposed Channel Selection Scheme}
\label{scc:fairness}
\begin{figure}[hbt]
\subfigure[Fairness of the proposed channel selection scheme.]
{\includegraphics[width=0.49\textwidth]{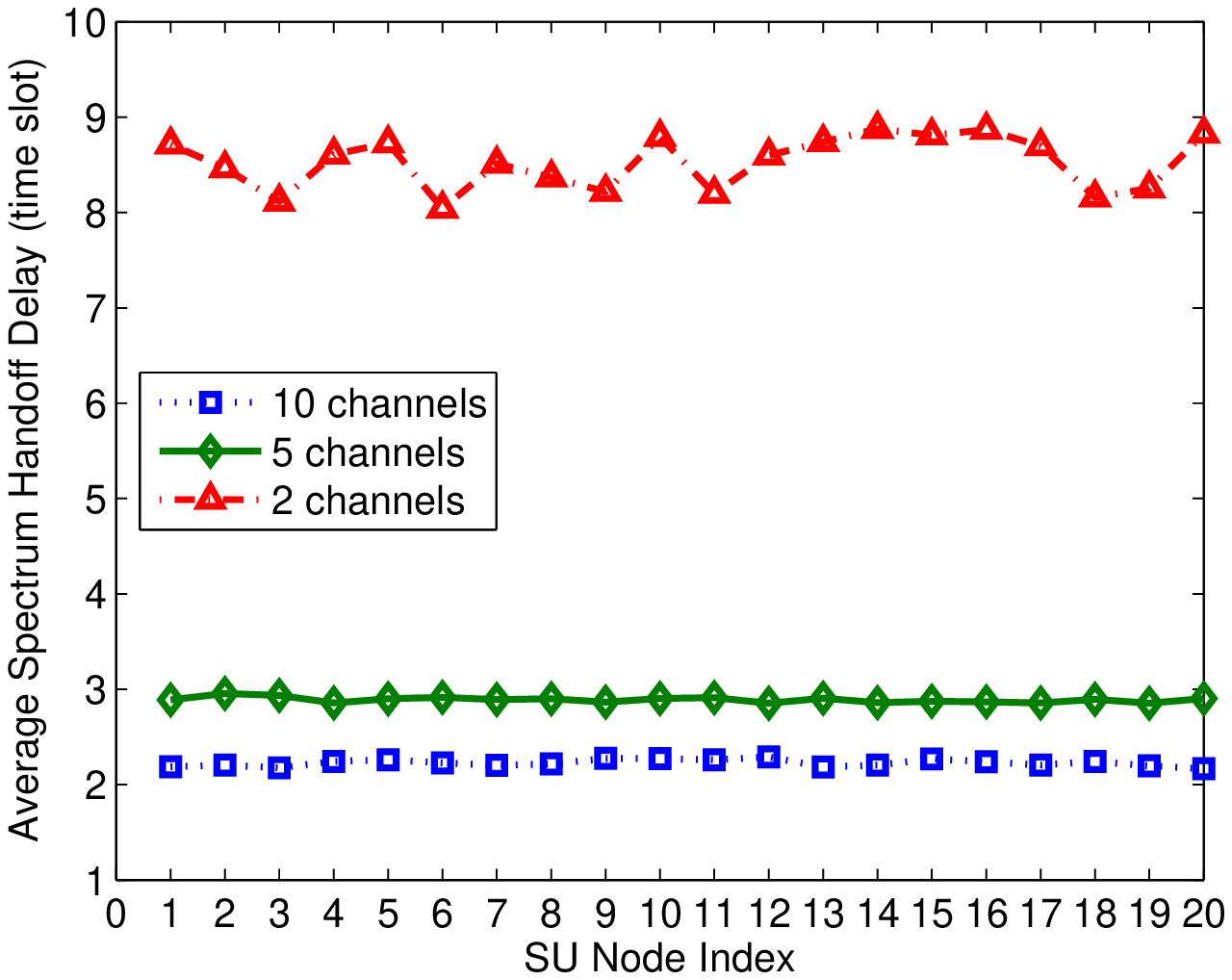}}\label{fig:fairness}
\subfigure[Scalability of the proposed channel selection scheme.]
{\includegraphics[width=0.49\textwidth]{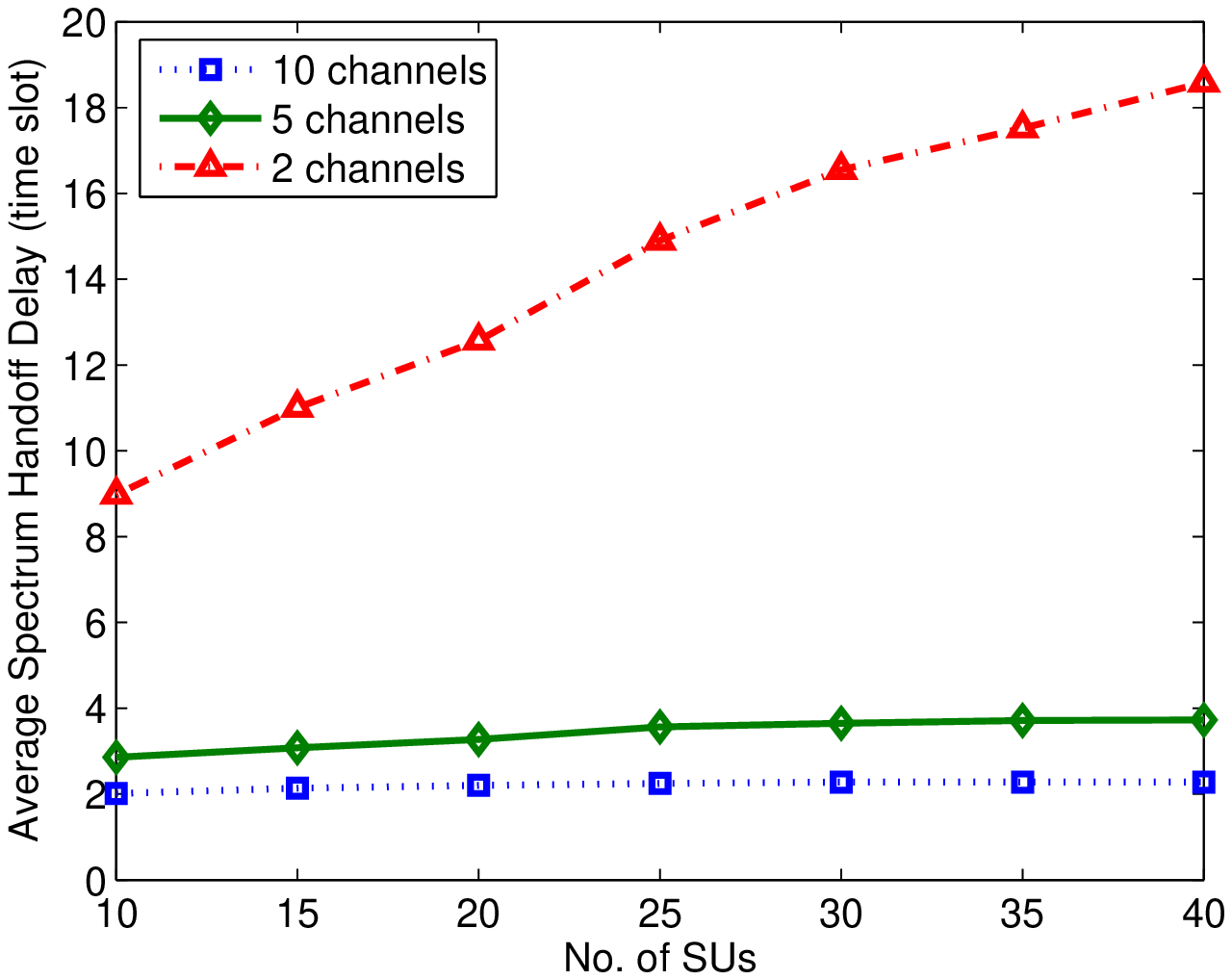}}\label{fig:scalability}
\label{fig:selection2}
\caption{Fairness and scalability of the proposed channel selection scheme.}
\end{figure}

The above procedure shows that our proposed channel selection scheme can avoid collisions among SUs and it is a fully distributed algorithm. In addition, from the above discussion, we observe that an important feature of the proposed distributed channel selection scheme is fairness. Unlike the previous definition of fairness as equal channel capacity for every user \cite{Zheng05}, in this paper, we define fairness as equal average handoff delay for every SU. This is because that, from the network performance point of view, handoff delay is the most significant metric to evaluate a spectrum handoff protocol. Thus, letting every SU have equal average handoff delay is fair. We define the spectrum handoff delay as the duration from the moment a SU starts to perform a spectrum handoff to the moment it resumes the data transmission. Fig. \ref{fig:fairness} shows the simulation result of the average handoff delay of the SUs when they use the proposed channel selection scheme under the single rendezvous coordination scheme. We deploy 20 SU nodes in the network with different arrival rate which is a uniform random variable in the range of $[0,500]$ (unit: packet/second). It is shown in the figure that SUs achieve approximately the same average spectrum handoff delay in the same scenario, which indicates that our proposed channel selection scheme is fair to all SUs.

On the other hand, for CR ad hoc networks where nodes membership may change over time, an important issue is the scalability of the proposed channel selection algorithm when the network size increases. Even though the number of SUs in a network may vary, as illustrated in Algorithm \ref{al:targetchannel}, only those SUs who are involved in the spectrum handoff process at the same time will activate the algorithm, which may not be a large number. In addition, from the number of broadcasted messages during the second step of the proposed channel selection scheme, our proposed channel selection algorithm will not result in excessive overhead when the network size increases. Because the number of channel information message updates affects the spectrum handoff delay (i.e., more channel information messages updated results in longer spectrum handoff delay), Fig. \ref{fig:scalability} shows the simulation result of the average spectrum handoff delay under different network sizes. It is shown that when the network size changes from 10 SUs to 40 SUs (i.e., the network size increases $300\%$), the spectrum handoff delay only increases $14.5\%$, $16\%$, and $105\%$ for the cases when the number of channels is 10, 5, and 2, respectively.

\section{Performance Evaluation of the Proposed Proactive Spectrum Handoff Framework}
\label{sc:evaluation}
\subsection{Simulation Setup}
\label{setup}
In this section, we adjust the spectrum handoff criteria and policies proposed in Section \ref{ssc:swiching} to a time-slotted system and evaluate the performance of the proposed proactive spectrum handoff framework. In order for the system to be stable, we assume that the inter-arrival time of SU packets follow a biased geometric distribution, where the probability mass function (pmf) of the biased geometric distributed inter-arrival time is given by \cite{Gebali08}: 
\begin{equation}
p(N=n)=\left\{
\begin{array}{ll}
0 & n<a\\
x(1-x)^{(n-a)} &n\geq a,
\end{array}
\right.
\label{eq:interarrival2}
\end{equation}
where $n$ is the number of time slots between packet arrivals, $a\geq0$ represents the minimum number of time slots between two adjacent packets, and $x$ is the probability that a packet arrives during one time slot (i.e., $x$ is the normalized arrival rate of data packets, that is, $x=\lambda\beta$, where $\lambda$ is the arrival rate in terms of packet/second). Based on this model, if we set $a$ as the packet length, then a new packet will not be generated until the previous packet finishes its transmission. 

\begin{figure}[hbt!]
\centering
\subfigure[No PU packet arrives between slot 1 and n.]
{\includegraphics[width=0.74\textwidth]{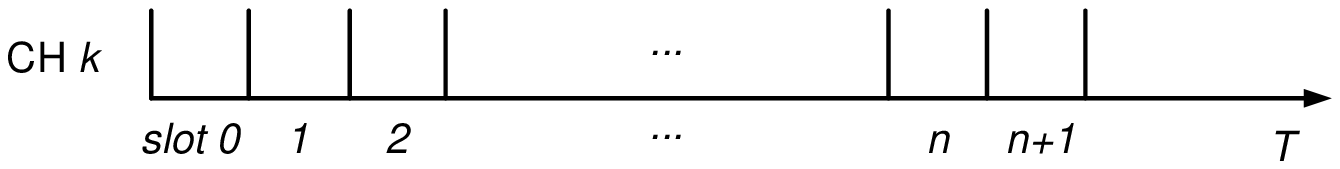}\label{fig:slotted1}}
\subfigure[Only one PU packet arrives between slot 1 and n.]
{\includegraphics[width=0.77\textwidth]{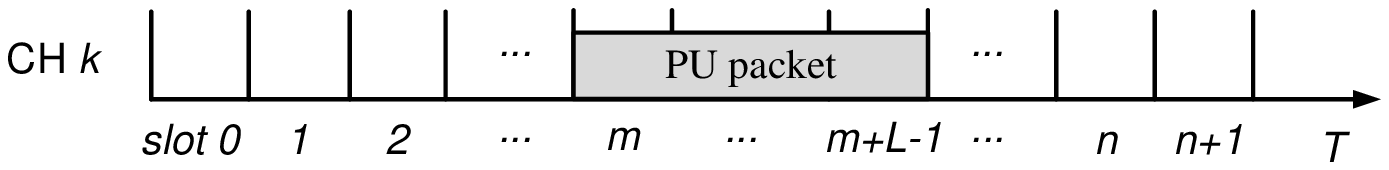}\label{fig:slotted2}}
\subfigure[$h$ PU packet arrives between slot 1 and n.]
{\includegraphics[width=0.9\textwidth]{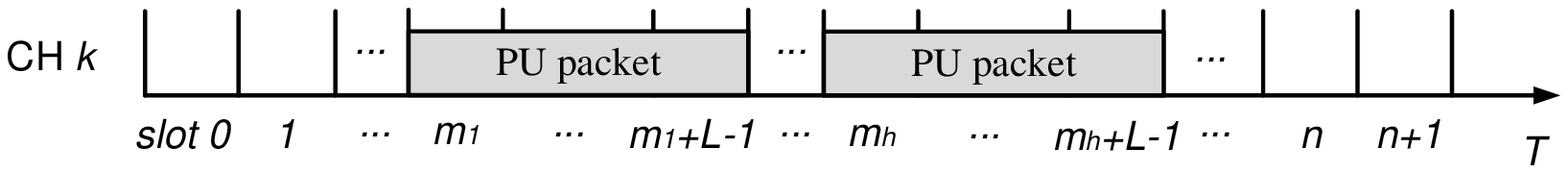}\label{fig:slotted3}}
\caption{The PU activity on channel $k$.}
\label{fig:slottedprediction}
\end{figure}

Accordingly, we modify the prediction criteria proposed in Section \ref{sc:protocol} based on the biased geometric distributed inter-arrival time model. Denote the starting slot of the prediction as slot 0 and the slot for prediction as slot $n$. As shown in Fig. \ref{fig:slotted1}, the probability that no PU arrival occurs between slot 1 and n and channel $k$ is idle at slot $n~ (n\geq1)$ is given by 
\begin{equation}
P_0=1-\sum_{i=1}^nx(1-x)^{(i-1)},
\label{eq:idle_2}
\end{equation}
where $x$ is the normalized arrival rate. As shown in Fig. \ref{fig:slotted2}, the probability that only one PU packet arrives between slot 1 and $n~(n>L)$ and channel $k$ is idle at slot $n$ is  
\begin{equation}
P_1=\sum_{m=1}^{n-L}\left[1-\sum_{i=1}^{n-m-L+1}x(1-x)^{(i-1)}\right]x(1-x)^{(m-1)},
\label{eq:idle1}
\end{equation}
where $m$ is the time slot at which a PU transmission starts and $L$ is the length of a PU packet. Similarly, in Fig. \ref{fig:slotted3}, $m_i$ denotes the time slot at which the $i$-th PU transmission starts. Thus, the probability that $h$ PU packets arrives ($h\in[1,U]$), where $U$ is the maximum number of PU packets that could arrives between slot 1 and $n~(n>hL)$ and channel $k$ is idle at slot $n$ is
\begin{equation}
P_h=\sum_{m_h=h}^{n-hL}\left[1-\sum_{i=1}^{n-m_h-hL+1}x(1-x)^{(i-1)}\right]x^h(1-x)^{(m_h-h)}.
\label{eq:idleh}
\end{equation}
Therefore, the total probability that channel $k$ is idle at slot $n$ is obtained as follows:
\begin{equation}
\Pr(N_k(n)=0)=\sum_{i=0}^{U}P_i.
\label{eq:idle2}
\end{equation}

Secondly, due to the memoryless property of geometric distribution, the probability that the duration of the idleness is longer than $\eta$ slots on channel $k$ is given by
\begin{equation}
P(t_{k,off}>\eta)=1-\sum_{i=1}^\eta x(1-x)^{(i-1)}.
\label{eq:idleness}
\end{equation}
In this chapter, we exclude the effect of the channel switching delay (i.e., RF configuration delay), but it can be easily taken into account when necessary.

\subsection{The Proposed Proactive Spectrum Handoff Scheme}
\label{ssc:proactiveresults}
We first compare the proposed proactive spectrum handoff scheme with the reactive spectrum handoff approach. In the reactive spectrum handoff approach, a SU transmits a packet without predicting the availability of the current channel at the moment when a frame ends (i.e., using the policy in (\ref{eq:switch})). That is, a SU does not change the current channel by the end of a frame if the previous frame is successfully received. A spectrum handoff occurs only if the on-going transmission actually collides with a PU transmission and the collided SU frame needs to be retransmitted. 

In order to conduct a fair comparison, we assume that the channel prediction is a capability of SUs (i.e., SUs select candidate channels based on the policy in (\ref{eq:candidate}) in both schemes). Therefore, the only difference between the proposed proactive spectrum handoff scheme and the reactive spectrum handoff scheme is the mechanism to trigger the spectrum handoffs. In addition, in order to solely investigate the performance of the two spectrum handoff schemes, we adopt a general random channel selection scheme (i.e., a SU randomly selects a channel from its candidate channels) in both schemes. 

\begin{figure}[hbt!]
\centering
\subfigure[The SU packet arrival rate $\lambda_s$=5 packets/s.]
{\includegraphics[width=0.32\textwidth]{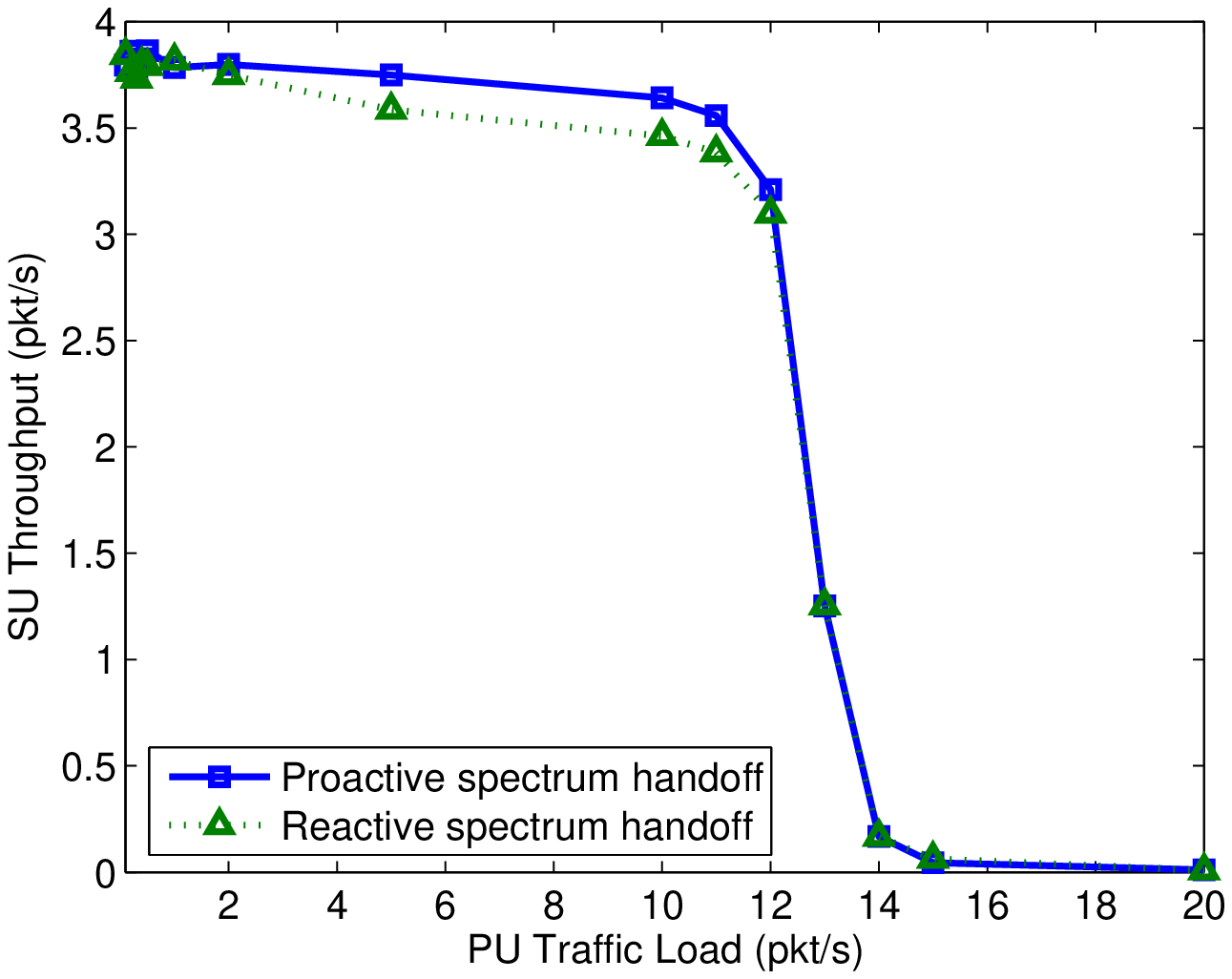}\label{fig:th5}}
\subfigure[The SU packet arrival rate $\lambda_s$=100 packets/s.]
{\includegraphics[width=0.32\textwidth]{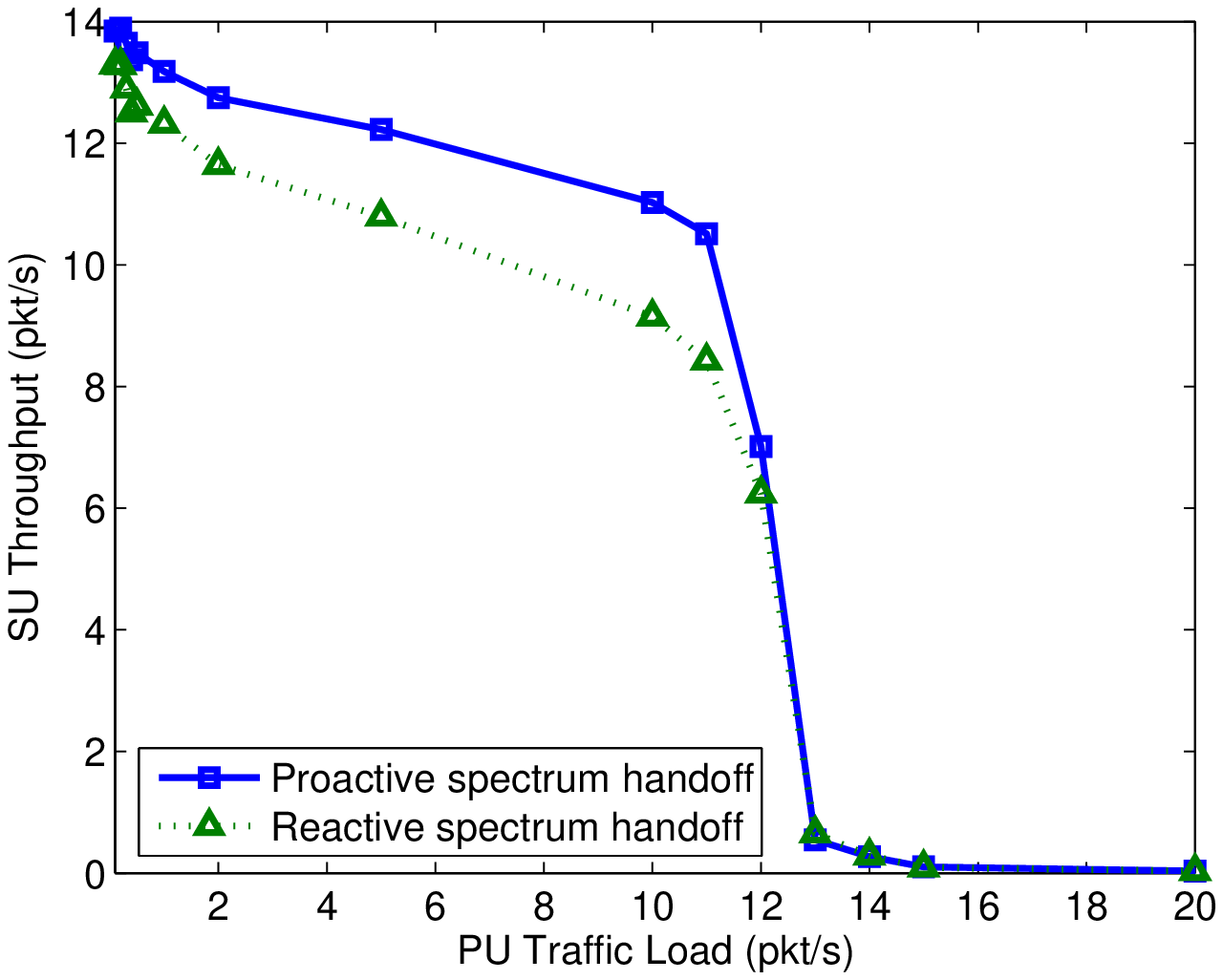}\label{fig:th100}}
\subfigure[The SU packet arrival rate $\lambda_s$=500 packets/s.]
{\includegraphics[width=0.32\textwidth]{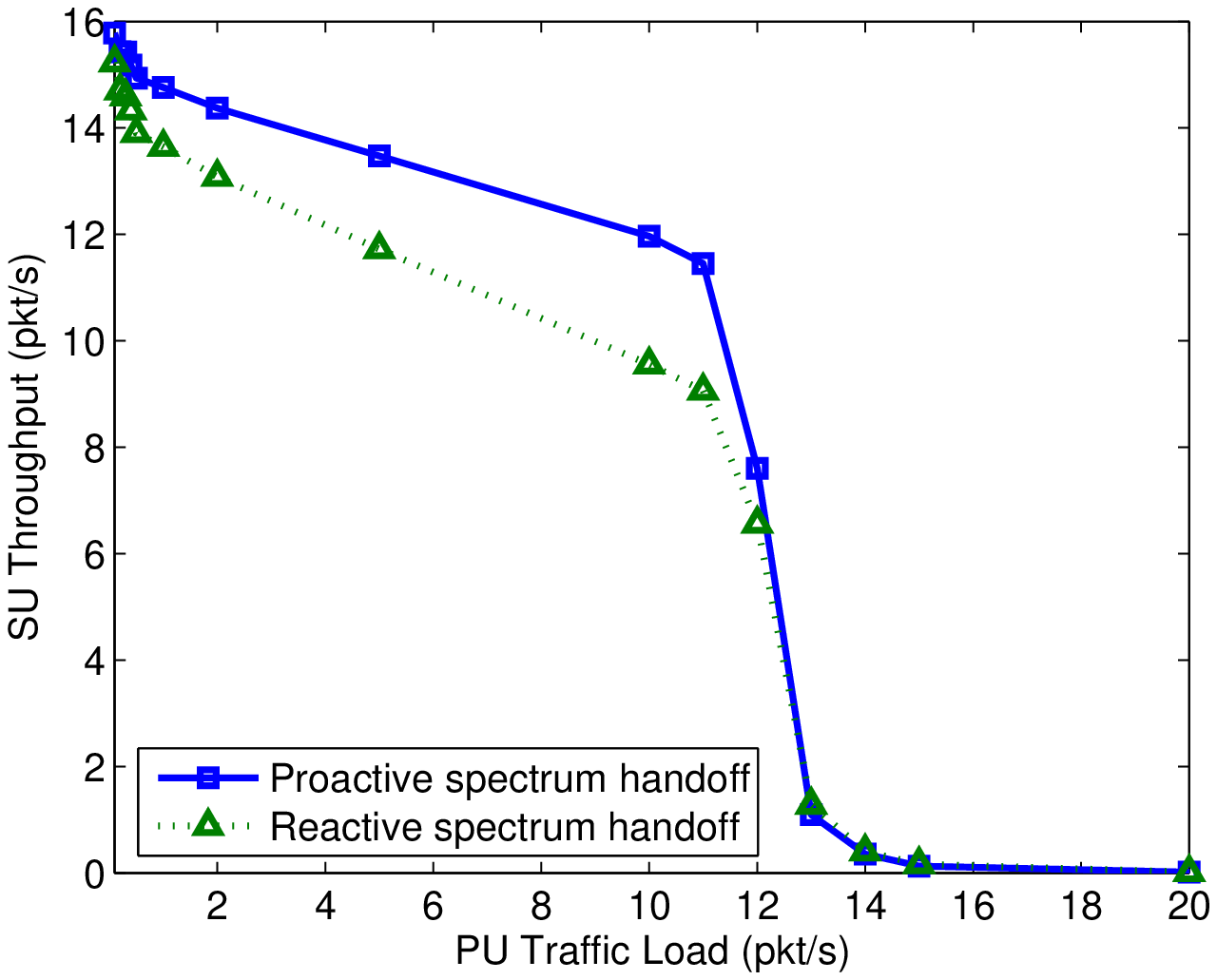}\label{fig:th500}}
\caption{Simulation results of SU throughput.}
\label{fig:thr1}
\end{figure}

\begin{figure}[hbt!]
\centering
\subfigure[The SU packet arrival rate $\lambda_s$=5 packets/s.]
{\includegraphics[width=0.32\textwidth]{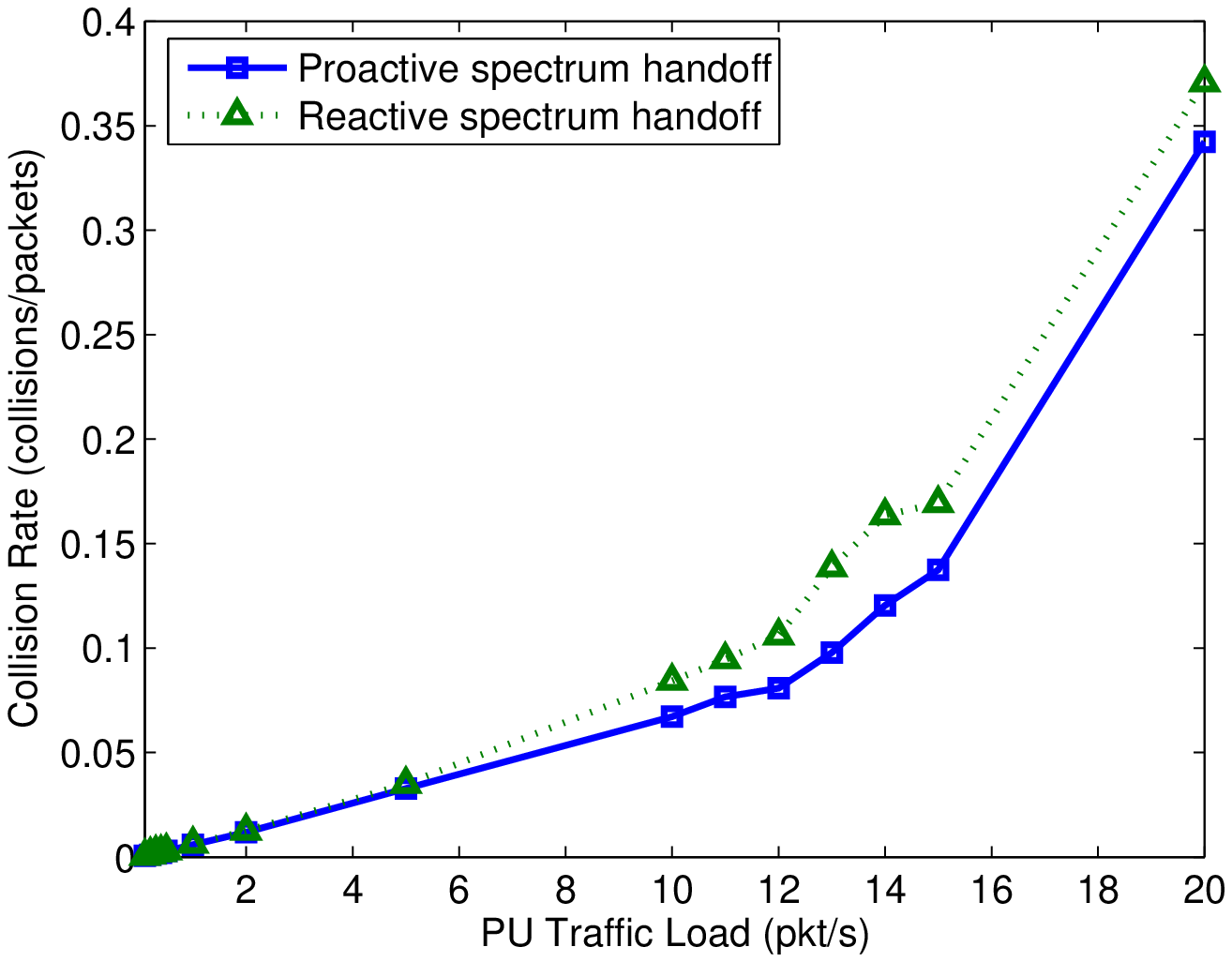}\label{fig:col5}}
\subfigure[The SU packet arrival rate $\lambda_s$=100 packets/s.]
{\includegraphics[width=0.32\textwidth]{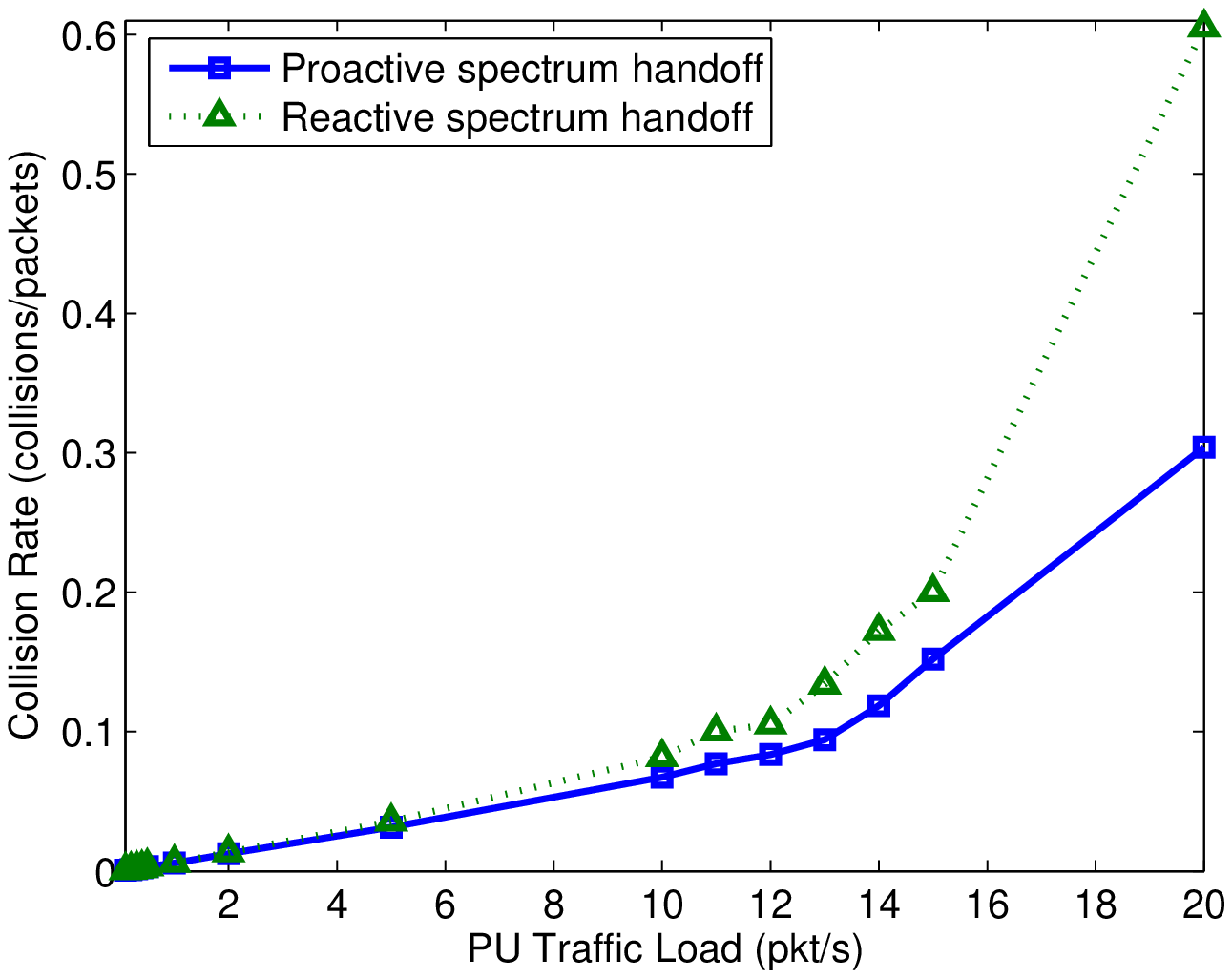}\label{fig:col100}}
\subfigure[The SU packet arrival rate $\lambda_s$=500 packets/s.]
{\includegraphics[width=0.32\textwidth]{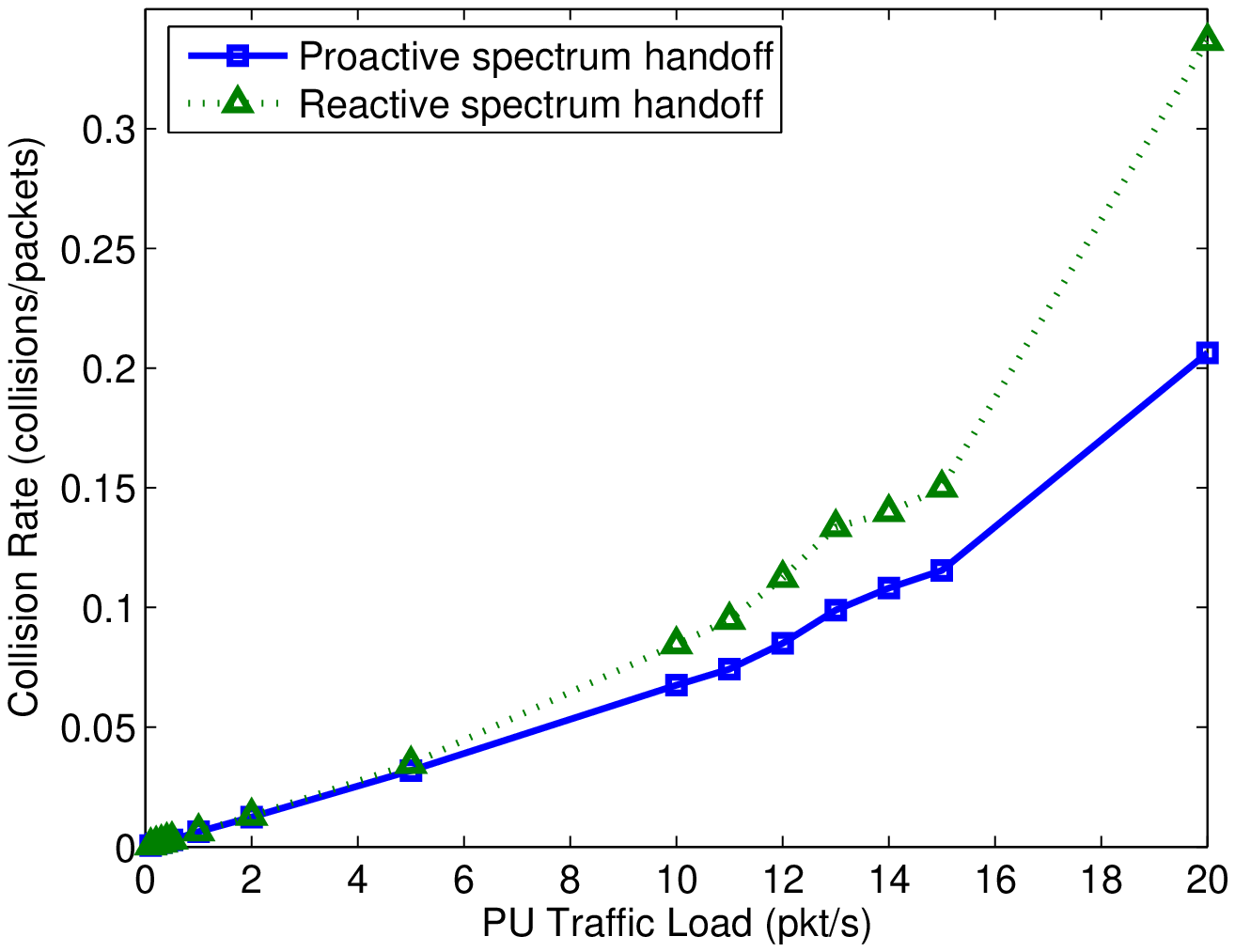}\label{fig:col500}}
\caption{Simulation results of collision rate.}
\label{fig:colrate}
\end{figure}

\begin{figure}[hbt!]
\centering
\subfigure[The SU packet arrival rate $\lambda_s$=5 packets/s.]
{\includegraphics[width=0.32\textwidth]{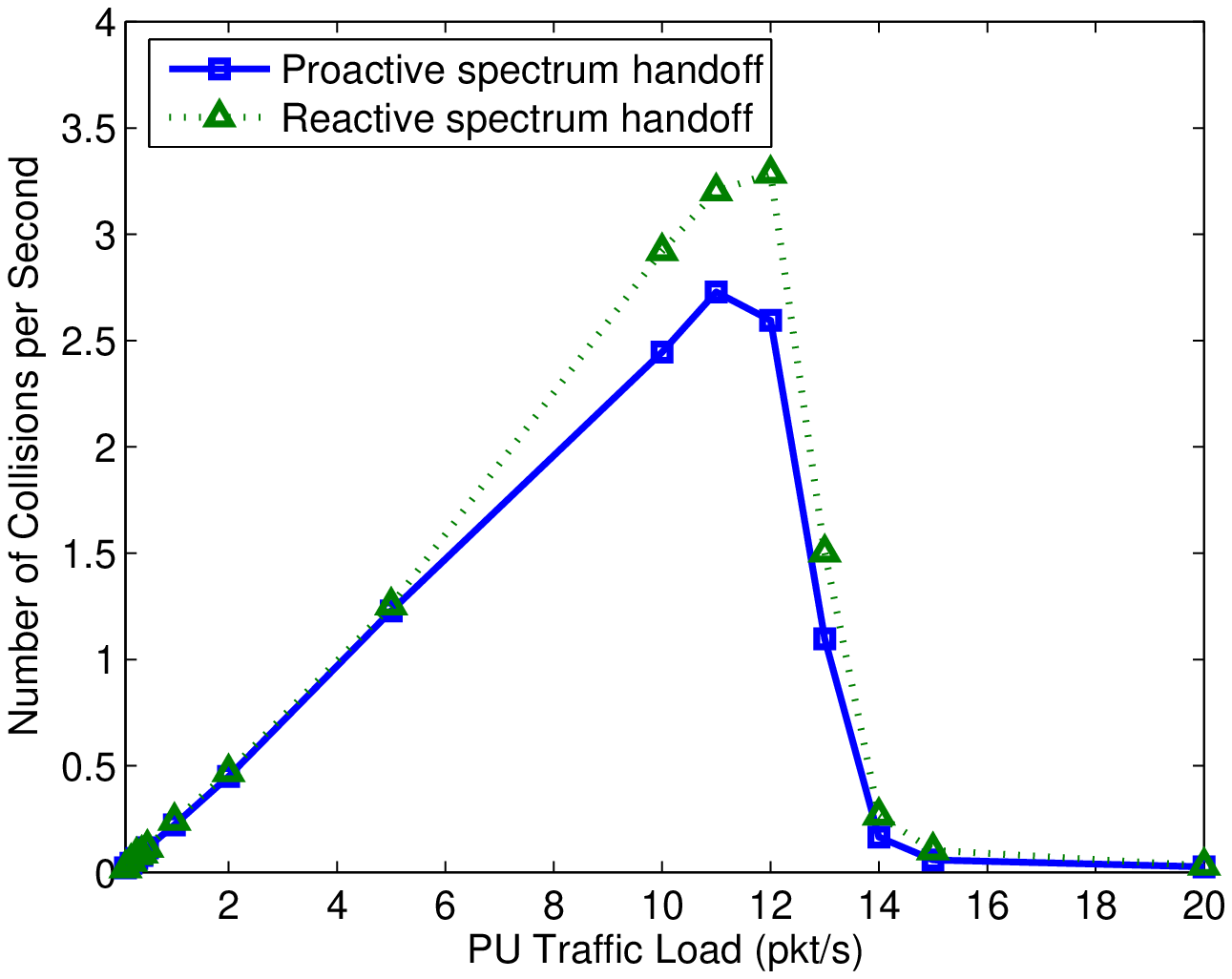}\label{fig:colnum5}}
\subfigure[The SU packet arrival rate $\lambda_s$=100 packets/s.]
{\includegraphics[width=0.32\textwidth]{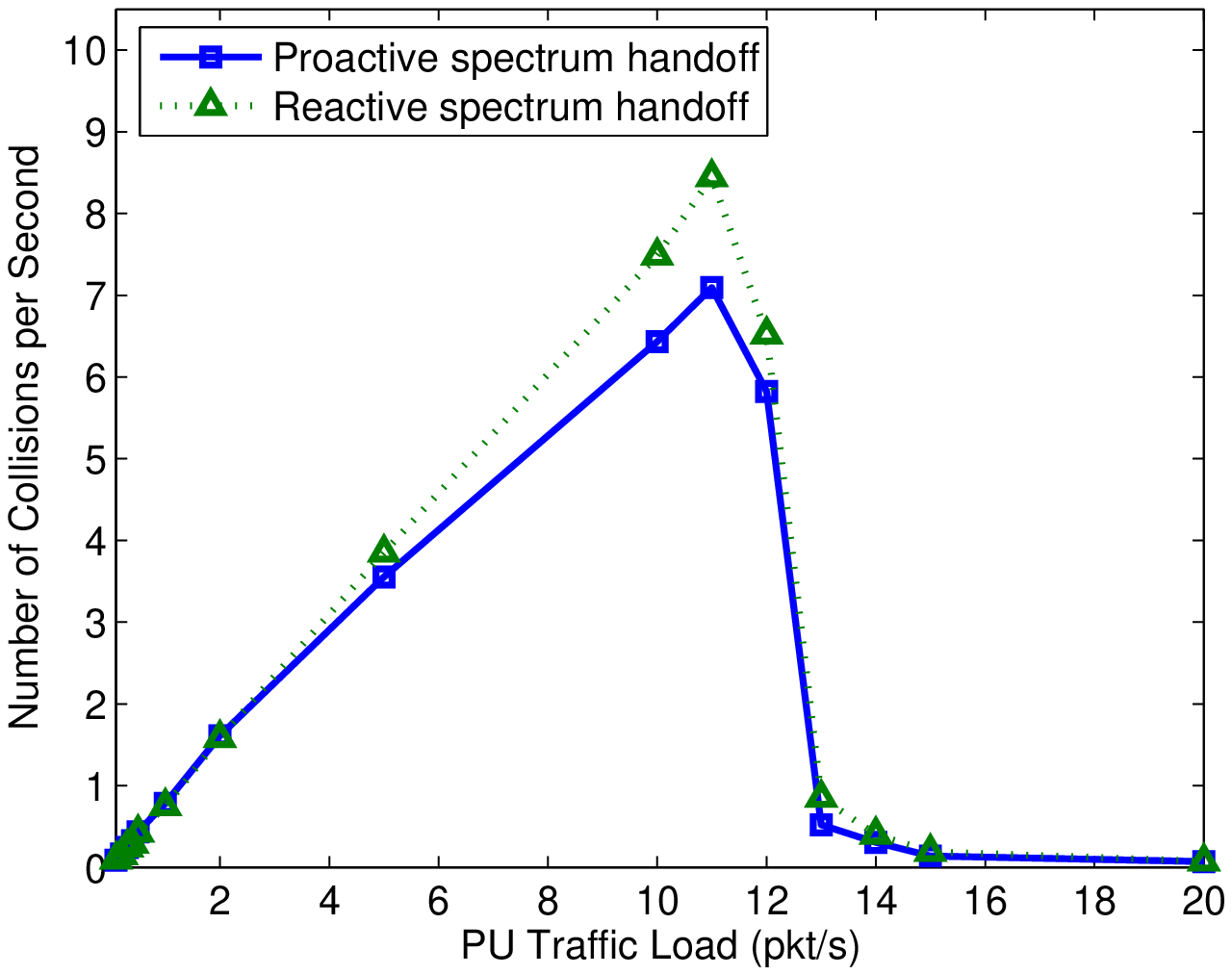}\label{fig:colnum100}}
\subfigure[The SU packet arrival rate $\lambda_s$=500 packets/s.]
{\includegraphics[width=0.32\textwidth]{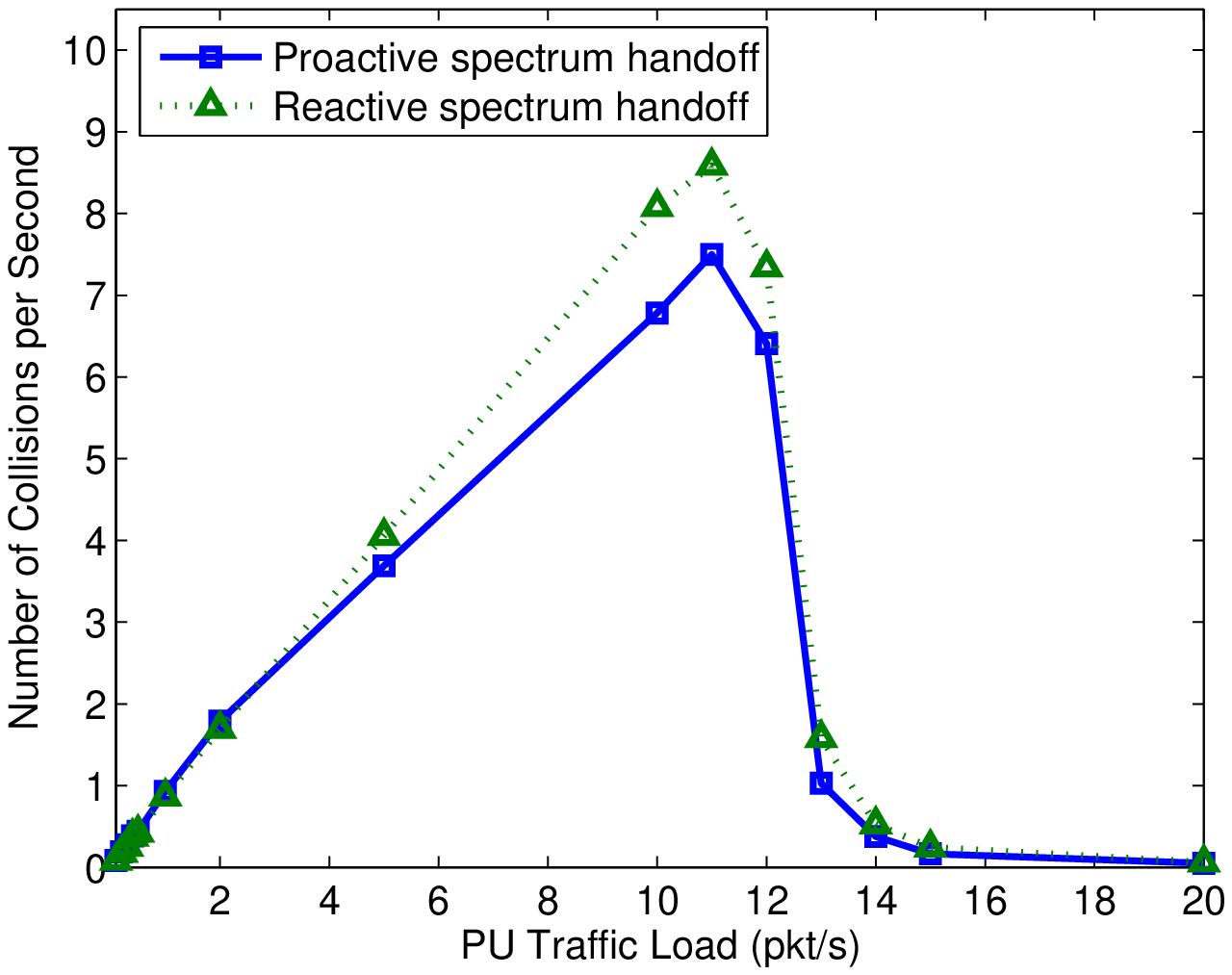}\label{fig:colnum500}}
\caption{Simulation results of number of collisions per second.}
\label{fig:colnum}
\end{figure}

Fig. \ref{fig:thr1} to Fig. \ref{fig:colnum} illustrate the performance results of the two spectrum handoff schemes under different SU and PU traffic load, when the network coordination scheme is the single rendezvous coordination scheme, where there are 10 SUs and 10 channels in the network. A SU using our proposed proactive spectrum handoff scheme will stop the data transmission on a channel which is likely to have a PU and switch to a channel which has less probability a PU appears. We choose the throughput of SUs, collision rate (i.e., the number of collisions between SUs and PUs per SU packet transmitted), and the number of collisions between SUs and PUs per second as the performance metrics.

Fig. \ref{fig:thr1} shows the SU throughput when SUs use different spectrum handoff schemes under varying SU and PU traffic load. It is shown that when both SU traffic and PU traffic are light (e.g., $\lambda_s$=5 packets/second and $\lambda_p$=0.5 packets/second), the SU throughput is similar in both schemes. This is because when the traffic is light, collisions between SUs and PUs are much fewer than the case when the traffic is heavy. SUs have less probability of retransmitting a packet for both cases, thus the performance differences between the proactive spectrum handoff scheme and the reactive spectrum handoff scheme are not very obvious. However, when the SU and PU traffic are heavy (e.g., $\lambda_s$=500 packets/second and $\lambda_p$=10 packets/second), the proactive spectrum handoff scheme outperforms the reactive scheme in terms of $30\%$ higher throughput. Fig. \ref{fig:colrate} and Fig. \ref{fig:colnum} show the collision rate and the number of collisions per second, respectively. From Fig. \ref{fig:colrate}, it is shown that collision rate increases as PU traffic load increases. In addition, proactive spectrum handoff always outperforms reactive spectrum handoff in terms of lower collision rate and fewer number of collisions per second.

\subsubsection{The Effect of the Number of SUs and PU channels}
\label{sssc:SU}
\begin{figure}[hbt!]
\centering
\subfigure[SU throughput.]
{\includegraphics[width=0.49\textwidth]{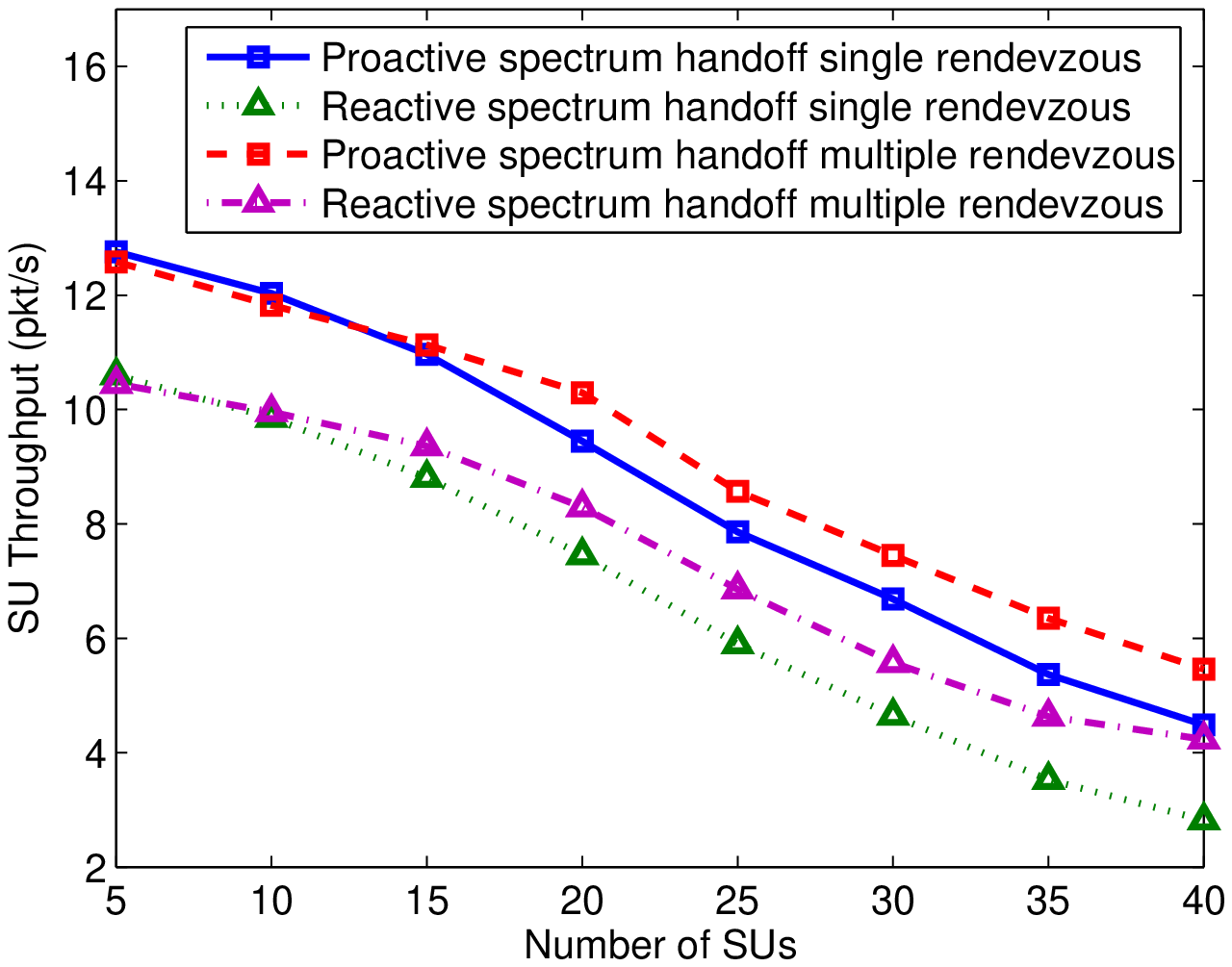}\label{fig:ssu}}
\subfigure[Collision rate.]
{\includegraphics[width=0.49\textwidth]{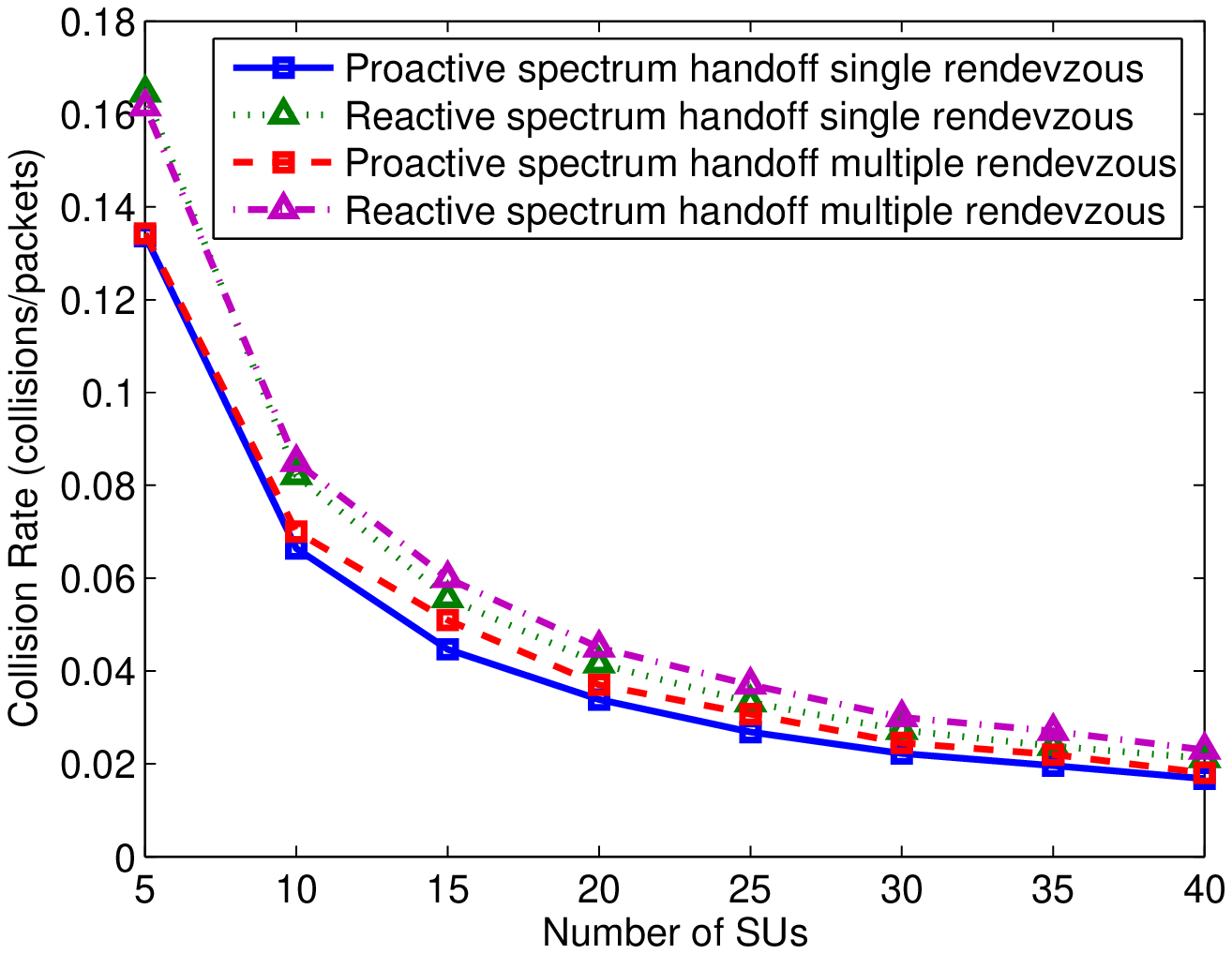}\label{fig:colsu}}
\caption{\small Performance comparison under different number of SUs.}
\label{fig:numsu}
\end{figure}

\begin{figure}[hbt!]
\centering
\subfigure[SU throughput.]
{\includegraphics[width=0.49\textwidth]{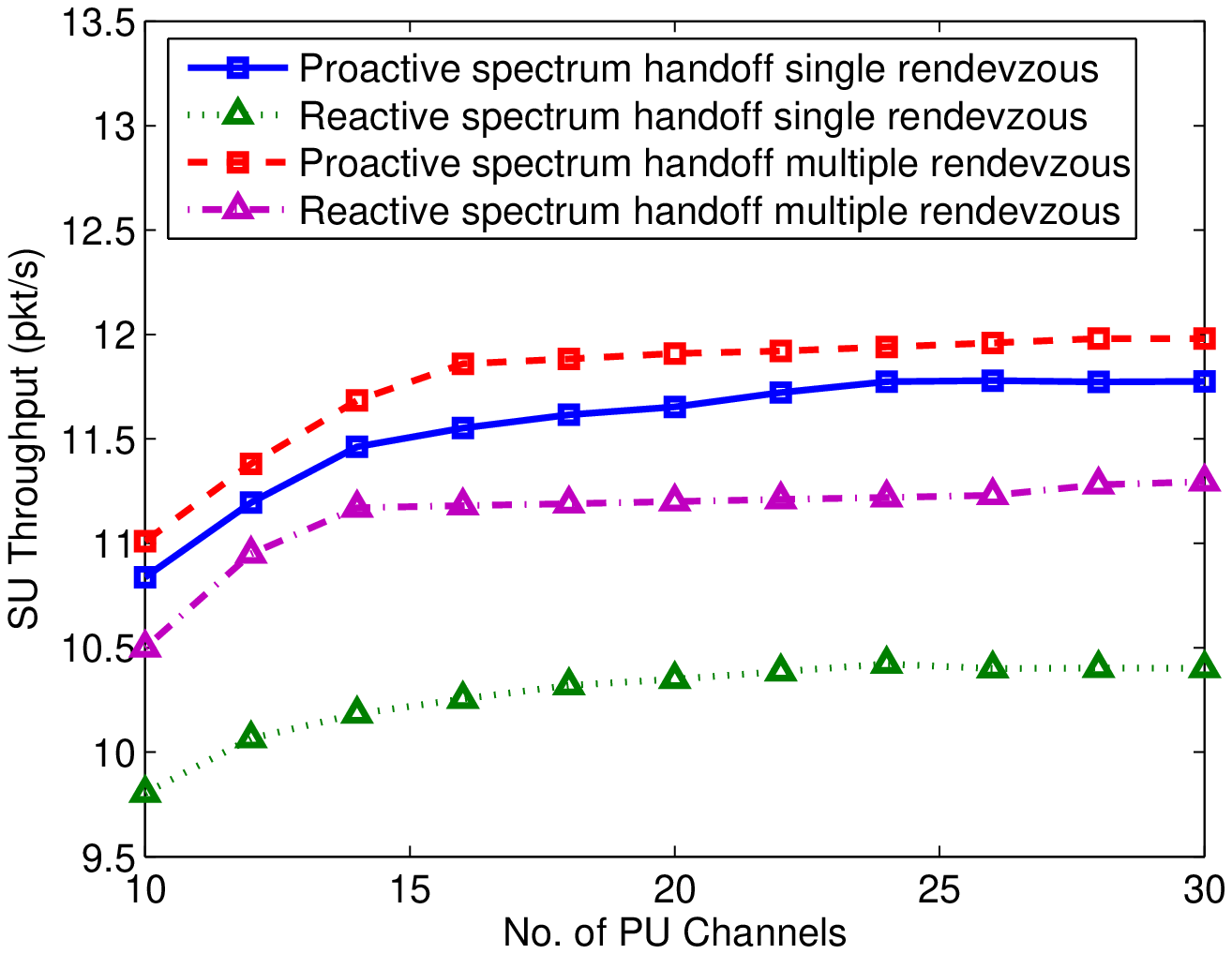}\label{fig:spu}}
\subfigure[Collision rate.]
{\includegraphics[width=0.49\textwidth]{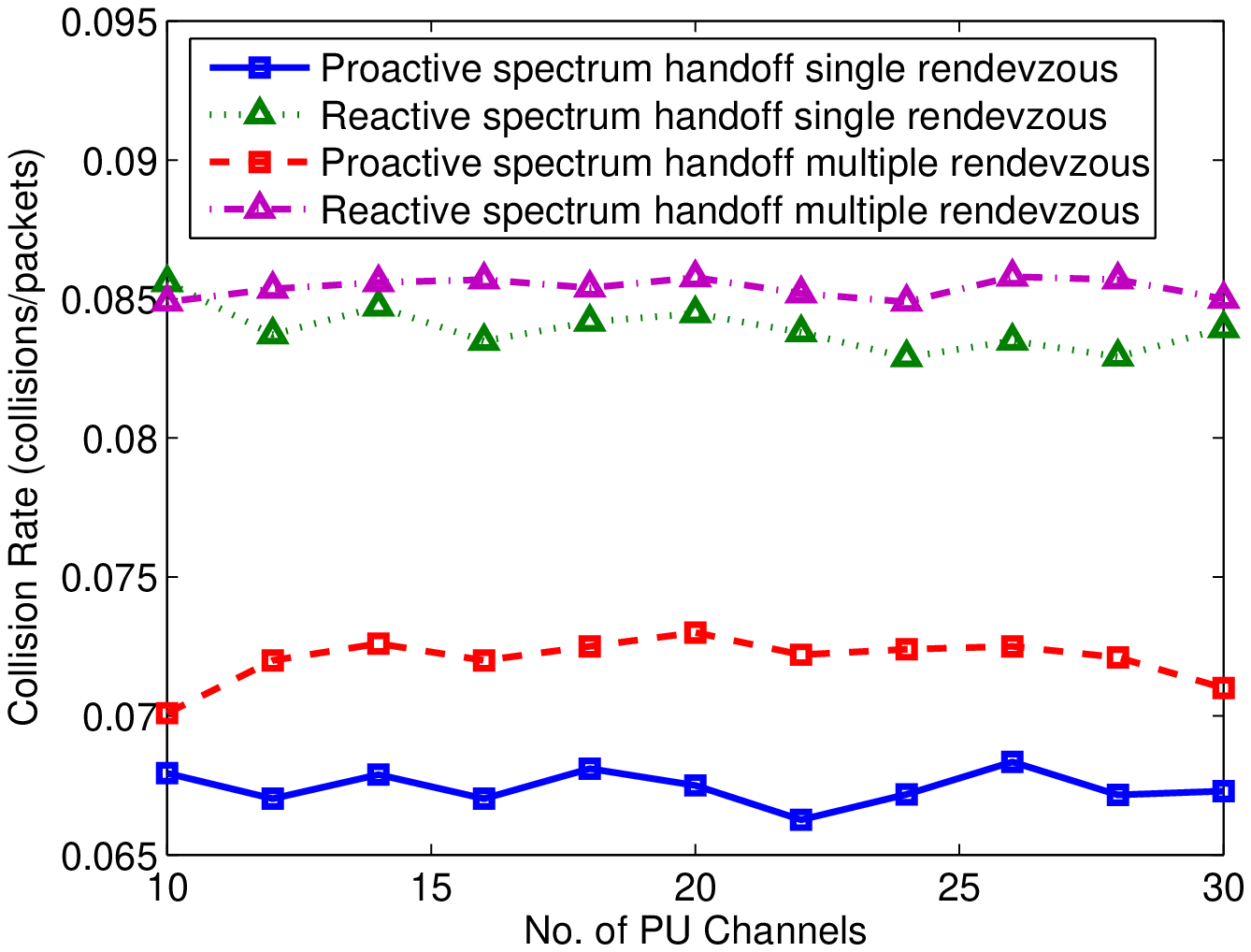}\label{fig:colpu}}
\caption{\small Performance comparison under different number of PU channels.}
\label{fig:numpu}
\end{figure}

Fig. \ref{fig:numsu} and \ref{fig:numpu} show the SU throughput and collision rate under varying number of SUs and PU channels, respectively. The results are generated in the scenario where the arrival rate of SU packets is saturated (i.e., $\lambda_s$=500 packets/second) and the arrival rate of PU packet is equal to 10 packets/second. In both figures, our proposed proactive spectrum handoff scheme outperforms the reactive spectrum handoff scheme in terms of higher SU throughput and lower collision rate. From Fig. \ref{fig:ssu} and Fig. \ref{fig:colsu}, it is shown that both the throughput and the collision rate of SU transmissions decreases as the number of SU increases. This is because that more SUs results in less opportunity of accessing the channel for each SU and causes higher probability of collisions among SUs when SUs initiate new transmissions or select channels when they perform spectrum handoffs. On the other hand, when the number of PU channels increases, the throughput of SUs first increases because more channels can be used for data transmissions. Then, the SU throughput becomes stable because increasing the number of channels does not help increasing the chance of data transmissions of SU packets after a certain threshold. The collision rate (i.e., the number of collisions between SUs and PUs per SU packet transmitted) remains relative stable to the change of the number of PU channels. Since in the multiple rendezvous coordination scheme, multiple pairs of SUs can use different channels to establish multiple links at the same time while only one pair is allowed to initiate a data transmission in the single rendezvous coordination scheme, the multiple rendezvous coordination scheme achieves higher SU throughput and lower collision rate than the single rendezvous coordination scheme, as shown in Fig. \ref{fig:numsu} and \ref{fig:numpu}.

\subsubsection{The Effect of the Length of SU and PU Packets}
\label{sssc:SUpacket}
\begin{figure}[hbt!]
\centering
\subfigure[SU throughput.]
{\includegraphics[width=0.49\textwidth]{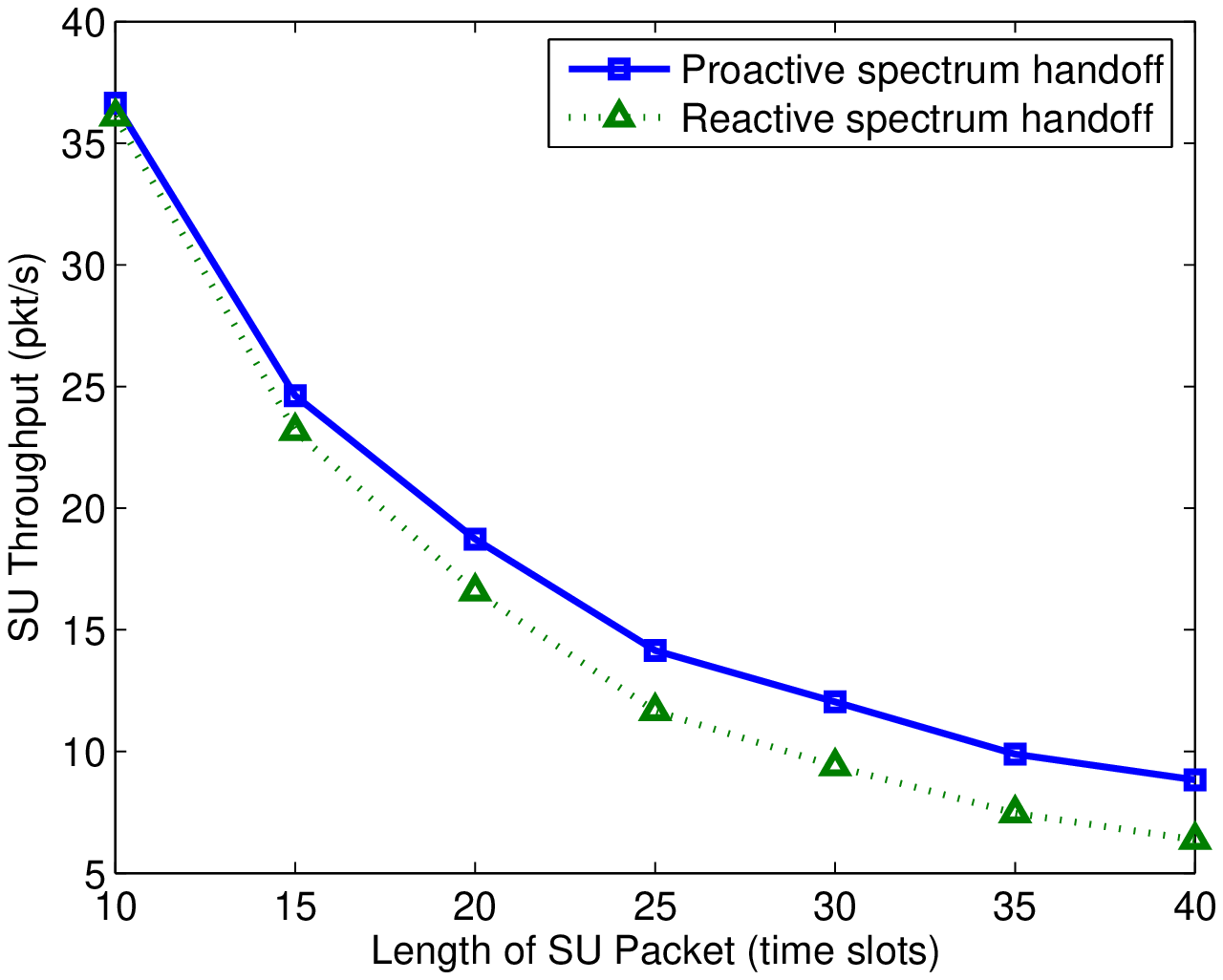}\label{fig:ssulen}}
\subfigure[Collision rate.]
{\includegraphics[width=0.49\textwidth]{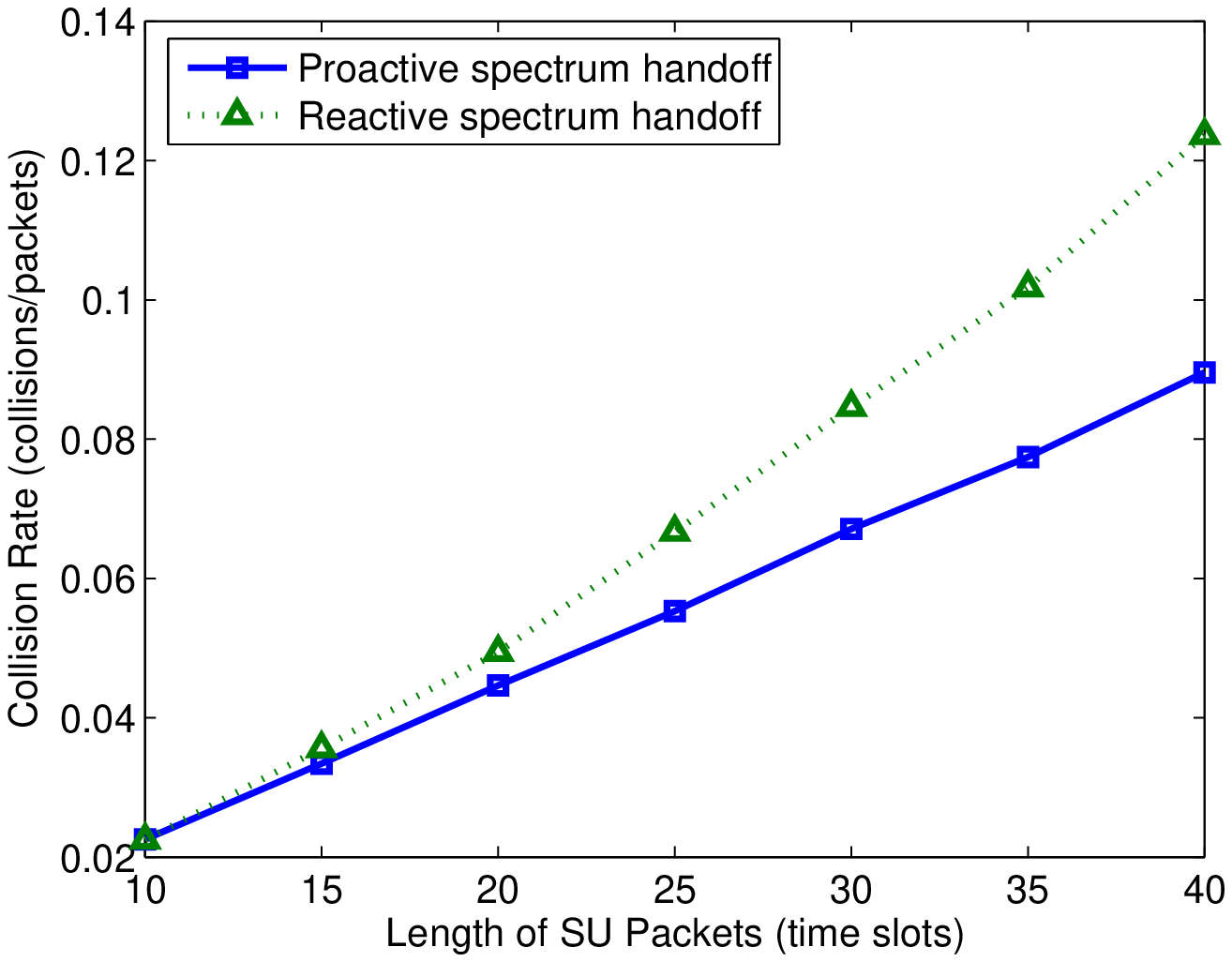}\label{fig:colsulen}}
\caption{\small Performance comparison under varying SU packet length.}
\label{fig:supacket}
\end{figure}

\begin{figure}[hbt!]
\centering
\subfigure[SU throughput.]
{\includegraphics[width=0.49\textwidth]{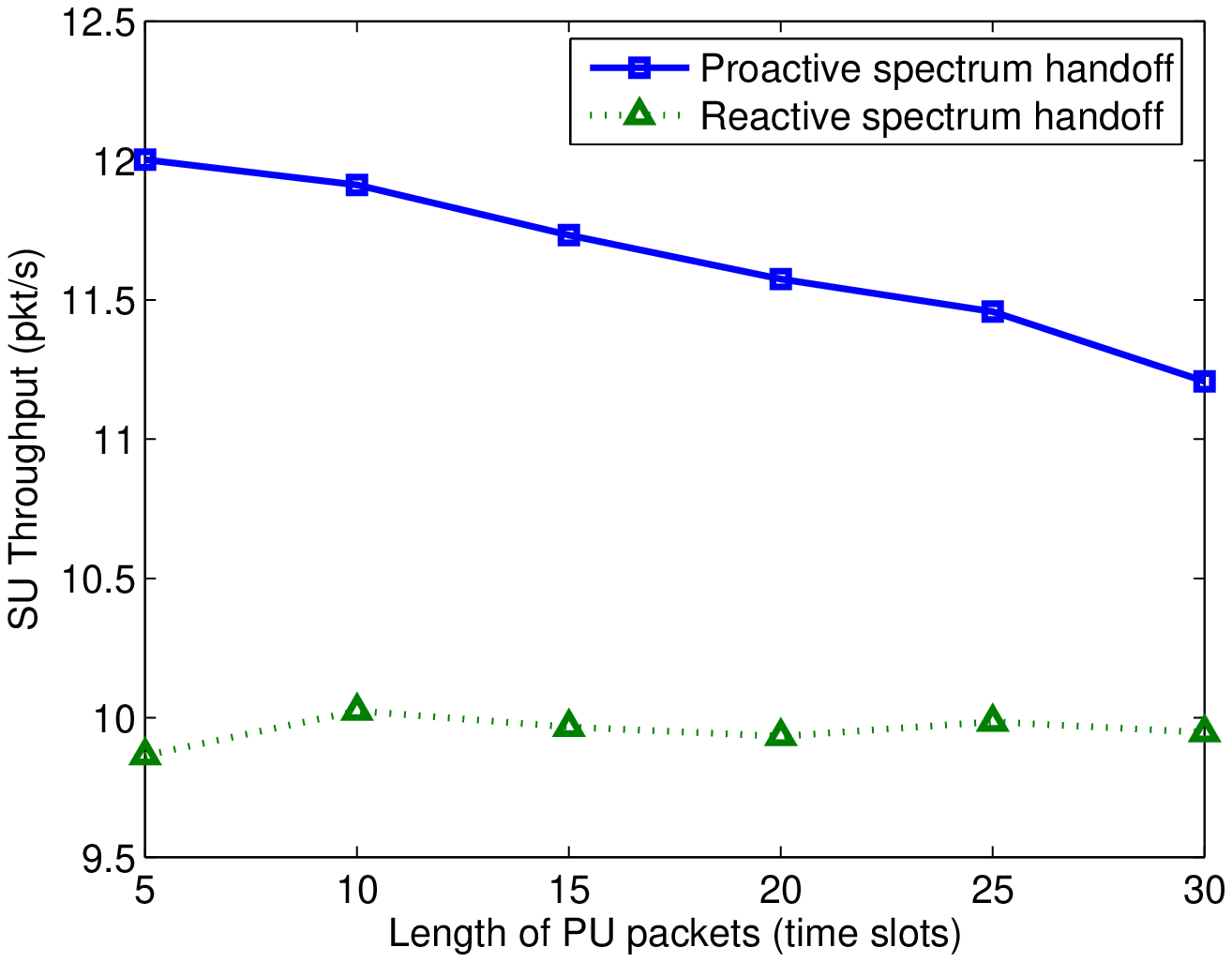}\label{fig:spulen}}
\subfigure[Collision rate.]
{\includegraphics[width=0.49\textwidth]{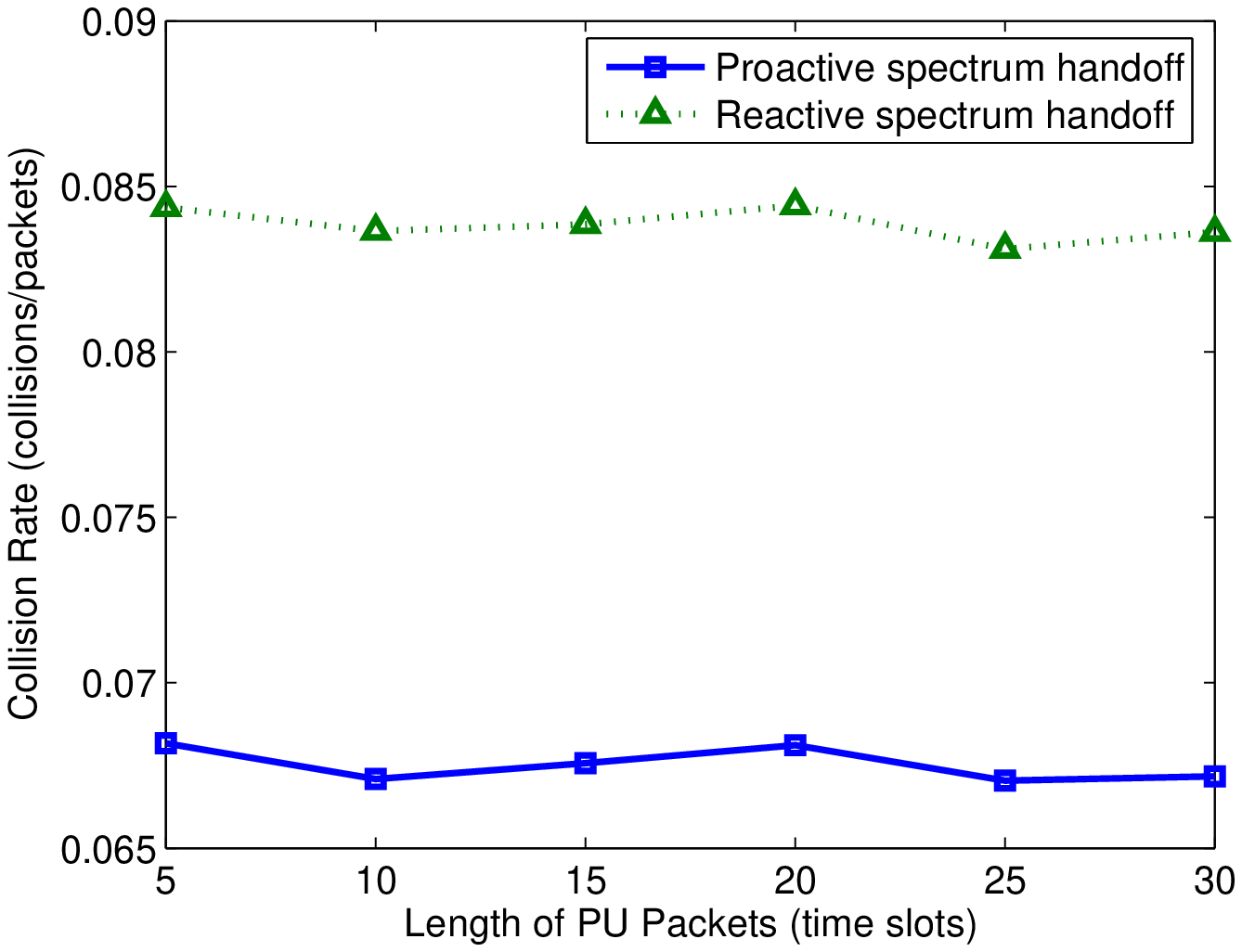}\label{fig:colpulen}}
\caption{\small Performance comparison under varying PU packet length.}
\label{fig:pupacket}
\end{figure}

Fig. \ref{fig:supacket} and \ref{fig:pupacket} show the SU throughput and collision rate under different lengths of SU and PU packets using the single rendezvous coordination scheme, respectively. It is shown in Fig. \ref{fig:ssulen} that when the length of SU packets increases, the throughput of SUs decreases because longer SU packet results in higher probability of collisions with PUs and leads to fewer SU packets transmitted during a certain amount of time. Therefore, it is illustrated in Fig. \ref{fig:colsulen} that the collision rate increases when the length of SU packets increases. On the other hand, the length of PU packets does not significantly affect the SU performance because we assume that once a SU frame collides with a PU packet, the whole frame needs to be retransmitted. Thus, the effect of the length of PU packets on SU performance is not significant.

\subsubsection{The Effect of Spectrum Sensing Errors}
\label{sssc:error}
\begin{figure}[hbt!]
\centering
\subfigure[SU throughput.]
{\includegraphics[width=0.49\textwidth]{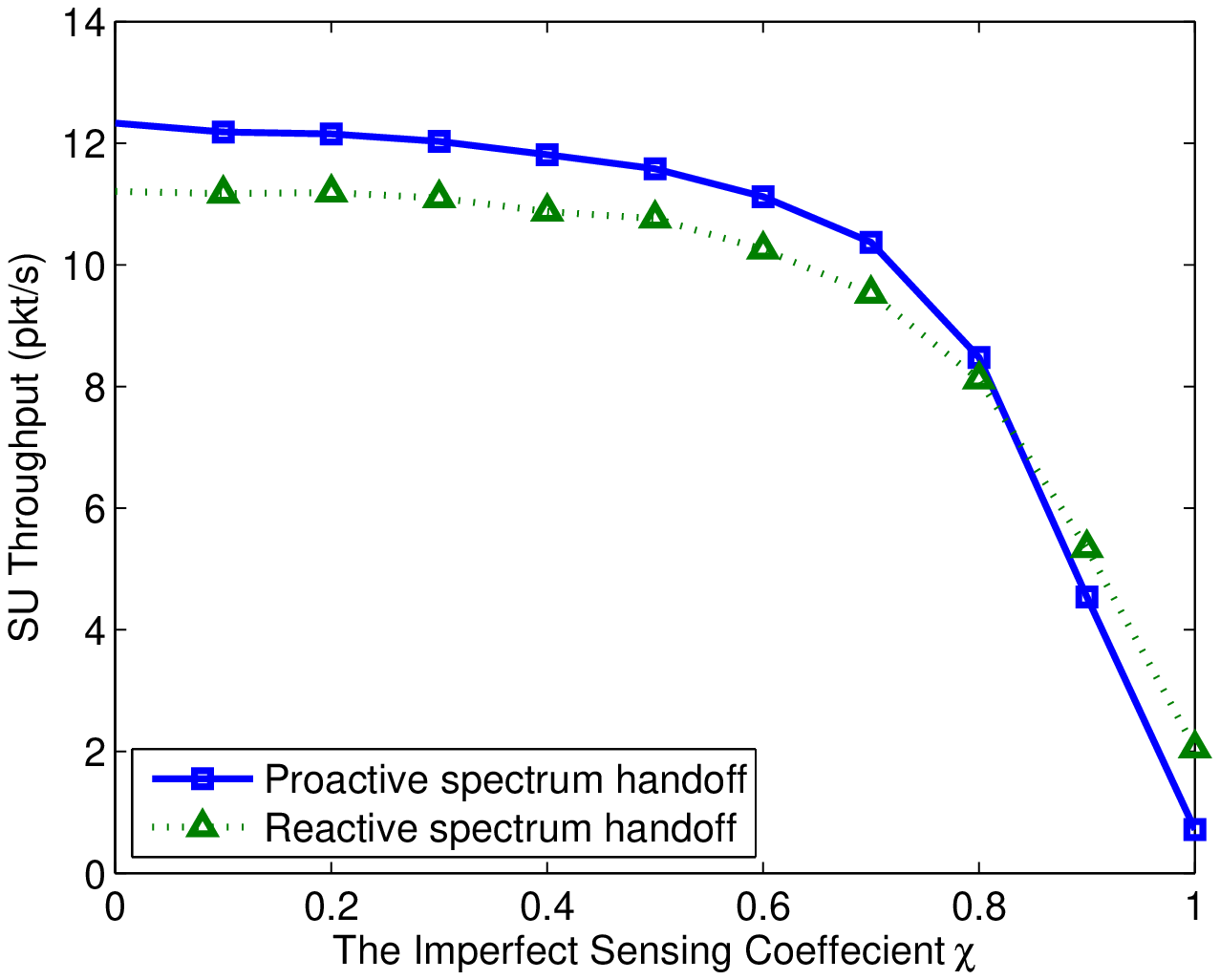}\label{fig:simp}}
\subfigure[Collision rate.]
{\includegraphics[width=0.49\textwidth]{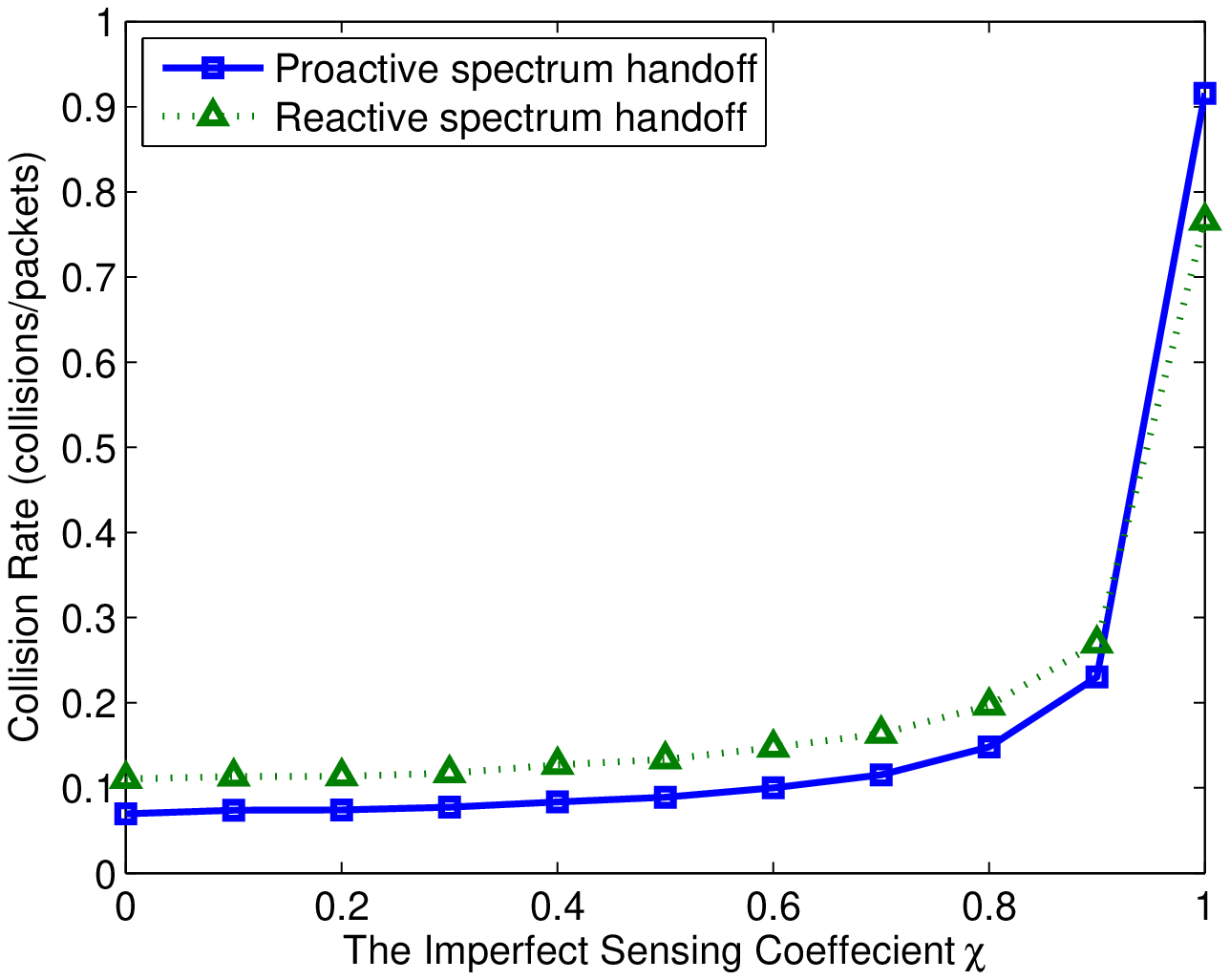}\label{fig:colimp}}
\caption{\small Performance comparison under imperfect spectrum sensing.}
\label{fig:imperfect}
\end{figure}

Fig. \ref{fig:imperfect} shows the effect of spectrum sensing errors on the performance of different spectrum handoff schemes using the single rendezvous coordination scheme. We use a coefficient $\chi$ to indicate the level of imperfect spectrum sensing, where $\chi \in [0,1]$ represents the probability that the result of spectrum sensing is wrong (the spectrum sensing errors include both miss detection and false alarm \cite{TangJ09}). When $\chi=0$, it means that the spectrum sensing is perfect and there is no error. Whereas when $\chi=1$, it means that the spectrum sensing is completely incorrect. It is shown in Fig. \ref{fig:imperfect} that the SU performance becomes worse as $\chi$ increases. However, the proposed proactive spectrum handoff scheme still outperforms the reactive spectrum handoff scheme in terms of higher throughput and lower collision rate.

\subsection{The Proposed Distributed Channel Selection Scheme}
\label{ssc:channelseletion}
To investigate the performance of the proposed channel selection scheme, we compare it with the following three different channel selection methods under the proposed proactive spectrum handoff scenario using the single rendezvous coordination scheme: \begin{itemize} 
\item \textit{Random channel selection}: A SU randomly chooses a channel from its predicted available channels.
\item \textit{Greedy channel selection}: In this method, only one pair of SUs is considered in the network. The SUs can obtain all the channel usage information and predict the service time on each channel. Thus, when a spectrum handoff occurs, a SU selects a pre-determined channel that leads to the minimum service time \cite{LCWang09}.
\item \textit{Local bargaining}: In this method, SUs form a local group to achieve a collision-free channel assignment. To make an agreement among SUs, a four-way handshake is needed between neighbors (i.e., request, acknowledgment, action, acknowledgment). Since one of the SUs is the initiating node which serves as a group header, the total number of control messages exchanged is $2N_{LB}$, where $N_{LB}$ is the number of SUs need to perform spectrum handoffs \cite{Zheng05}.
\end{itemize}
Since for channel selection schemes, reducing the number of collisions among SUs is the primary goal, we consider the SU throughput, average SU service time, collisions among SUs, and average spectrum handoff delay as the performance metrics.

\subsubsection{One-pair-SU Scenario}
\label{sssc:onepairscenario}
\begin{figure}[hbt]
\centering
\subfigure[SU throughput in a two-SU scenario.]
{\includegraphics[width=0.49\textwidth]{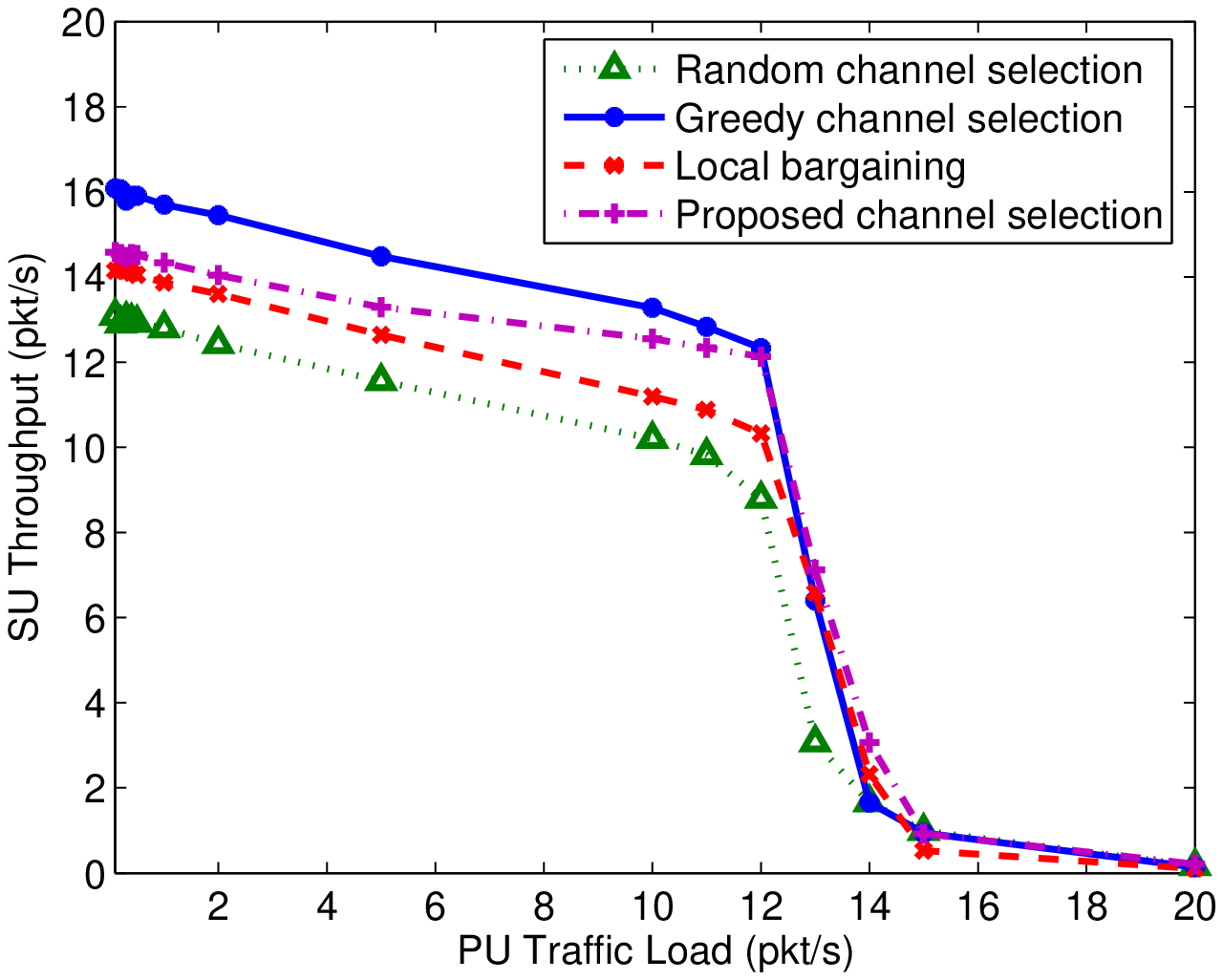}\label{fig:s_2nodes}}
\subfigure[SU average service time in a two-SU scenario.]
{\includegraphics[width=0.49\textwidth]{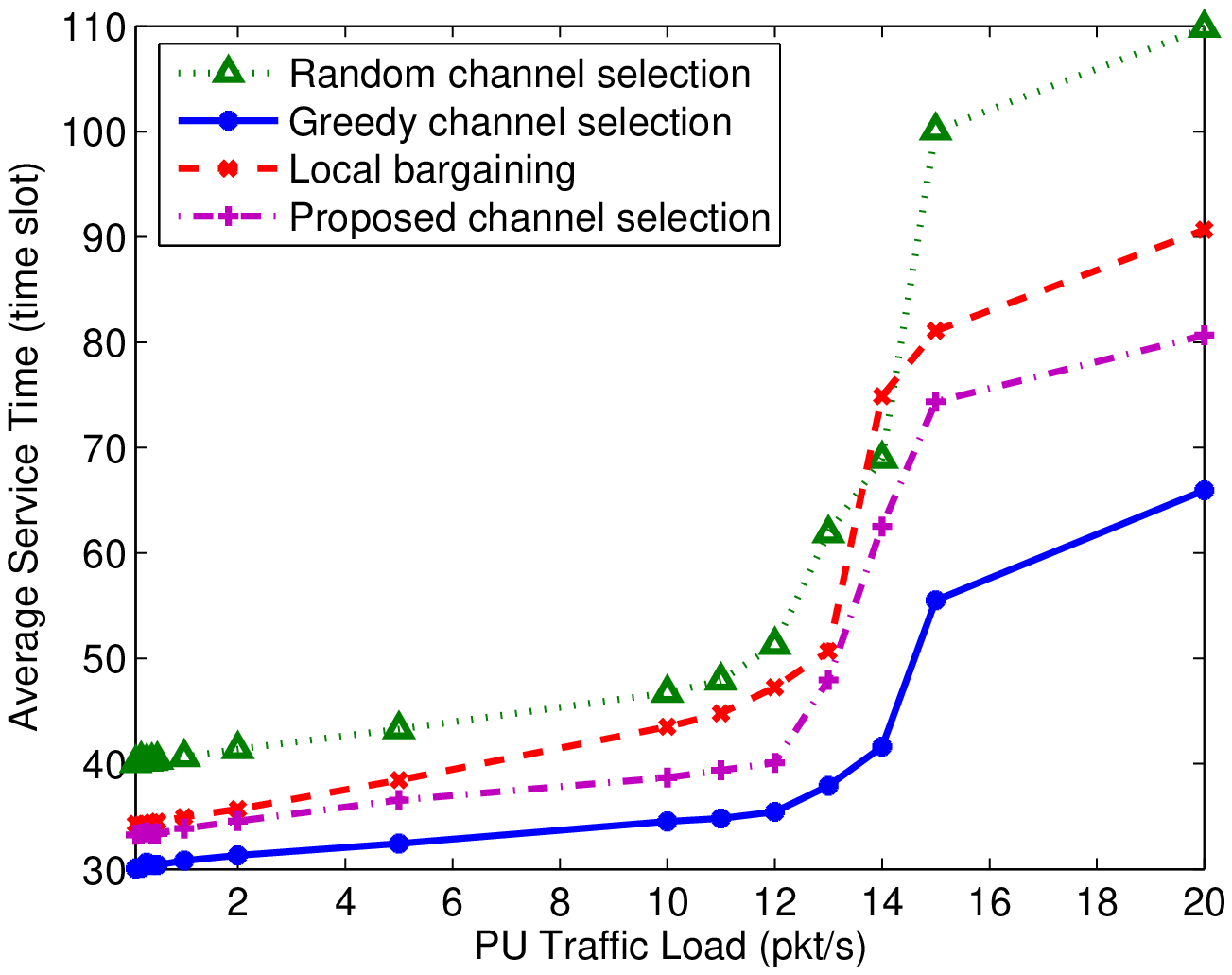}\label{fig:ast_2nodes}}
\caption{\small Performance of the channel selection schemes in a one-pair-SU scenario.}
\label{fig:channel_selection}
\end{figure}

Fig. \ref{fig:s_2nodes} and Fig. \ref{fig:ast_2nodes} show the SU throughput and the average service time of different channel selection schemes in a one-pair-SU scenario, respectively. Because only one pair of SUs exists in the network, there is no collision among SUs. Thus, in this scenario, the greedy channel selection scheme performs the best among all the schemes. This is because that the handoff target channel a SU transmitter selects is pre-determined based on channel observation history. Hence, no signaling message is needed between the SU transmitting pair. While in other schemes, the SU transmitter needs to inform the receiver about the newly selected channel. Thus, the throughput is lower and the average service time is longer than the greedy scheme. However, among the three schemes other than the greedy scheme, our proposed channel selection scheme has the best performance in terms of higher throughput and shorter total service time.

\subsubsection{Multiple-pair-SU Scenario}
\label{sssc:multipairscenario}
\begin{figure}[hbt]
\centering
\subfigure[SU throughput in a multi-SU scenario.]
{\includegraphics[width=0.49\textwidth]{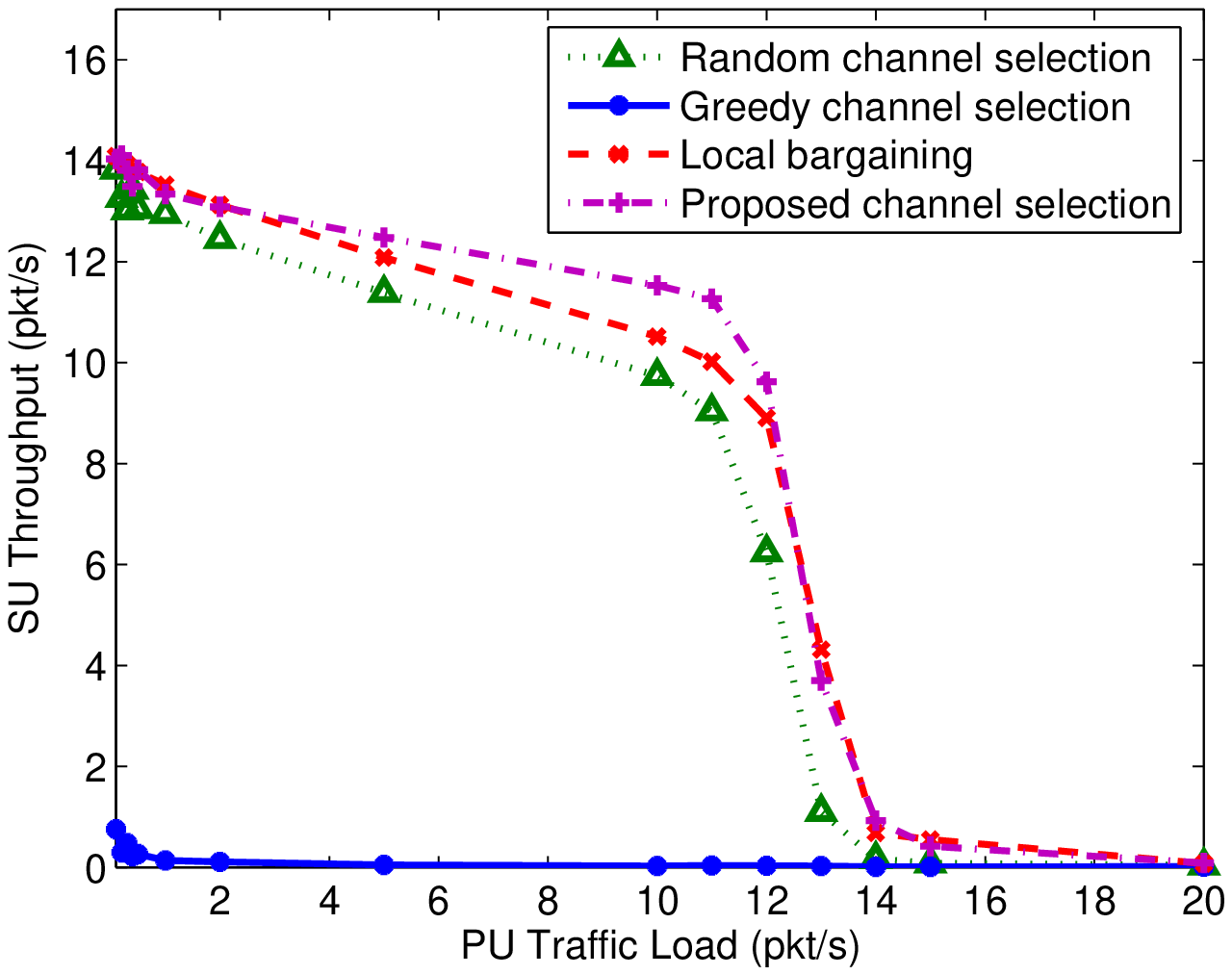}\label{fig:s_10nodes}}
\subfigure[SU average service time in a multi-SU scenario.]
{\includegraphics[width=0.49\textwidth]{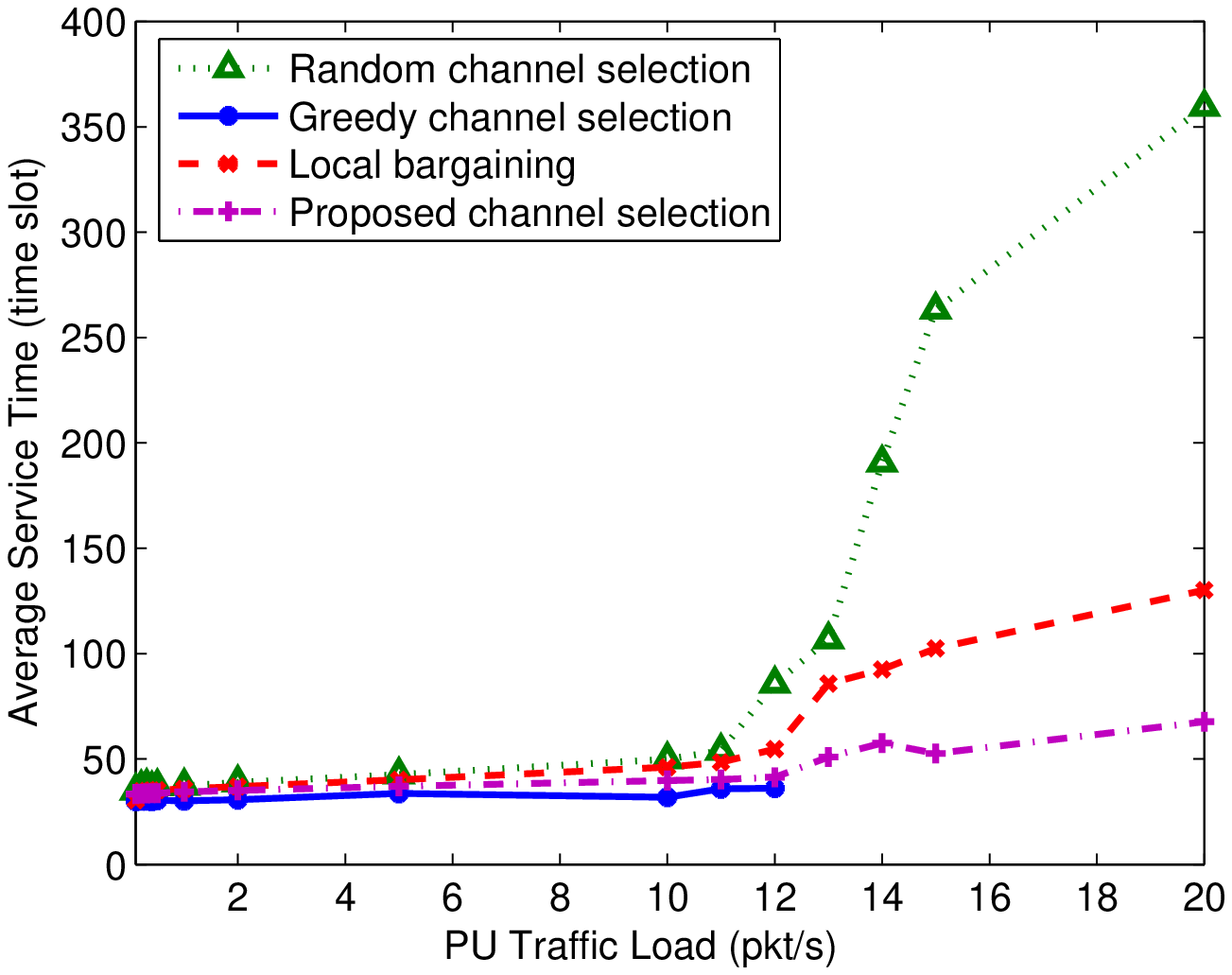}\label{fig:ast_10nodes}}
\caption{\small Performance of the channel selection schemes in a 10-pair-SU scenario.}
\label{fig:channel_selection_multi}
\end{figure}

Fig. \ref{fig:s_10nodes} and Fig. \ref{fig:ast_10nodes} show the SU throughput and the average service time of different channel selection schemes in a 10-pair-SU scenario, respectively. In the greedy channel selection method, all pairs of SUs always select the same pre-determined channel for spectrum handoffs. Therefore, the greedy method always leads to collisions among SUs. The throughput of SUs using the greedy method is almost zero. Because the proposed channel selection scheme can totally eliminate collisions among SUs, the throughput is higher and the average service time is shorter than the other channel selection schemes.

\begin{figure}[hbt]
\centering
\subfigure[SU throughput in a multi-SU scenario.]
{\includegraphics[width=0.49\textwidth]{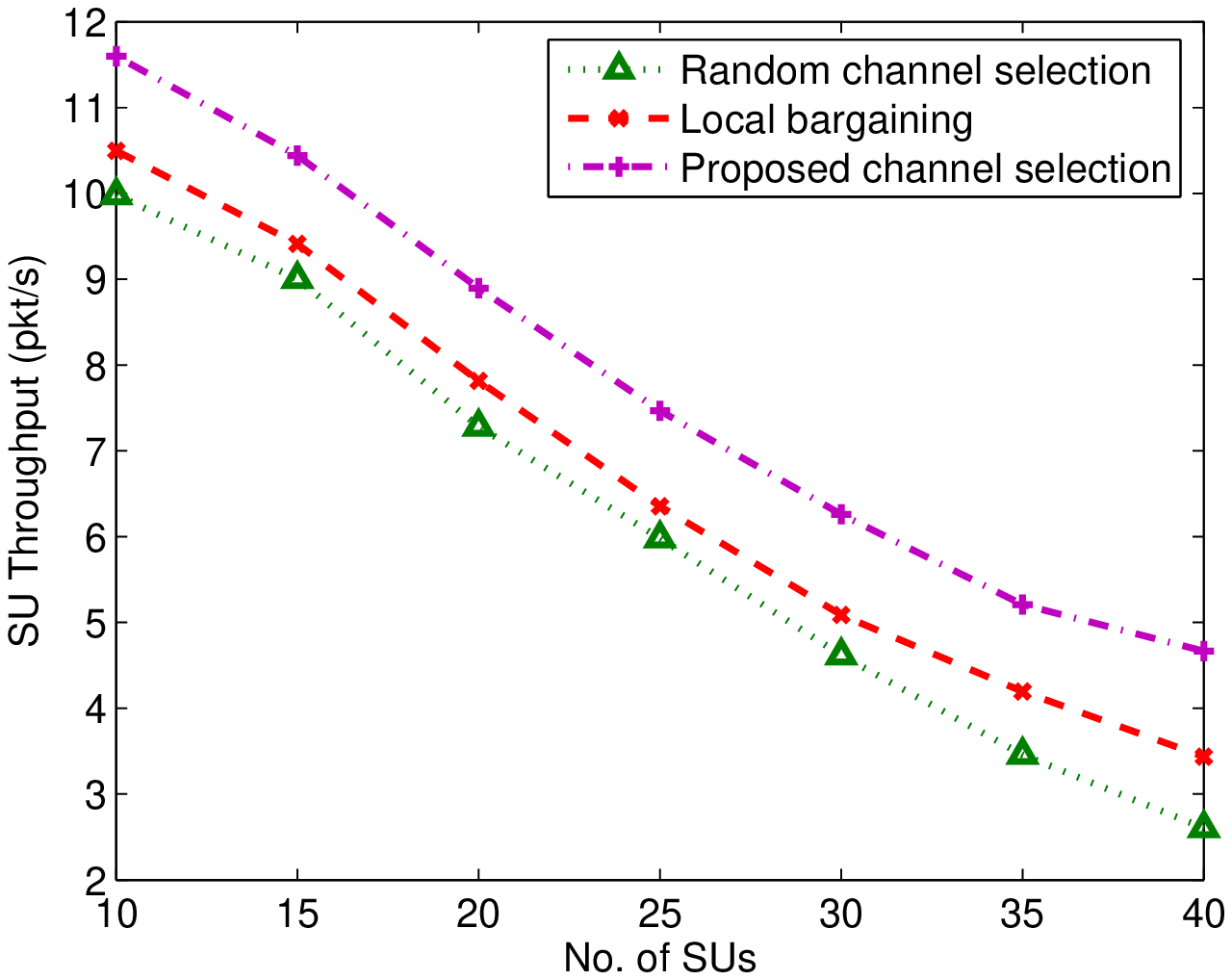}\label{fig:s_nodes}}
\subfigure[SU average service time in a multi-SU scenario.]
{\includegraphics[width=0.49\textwidth]{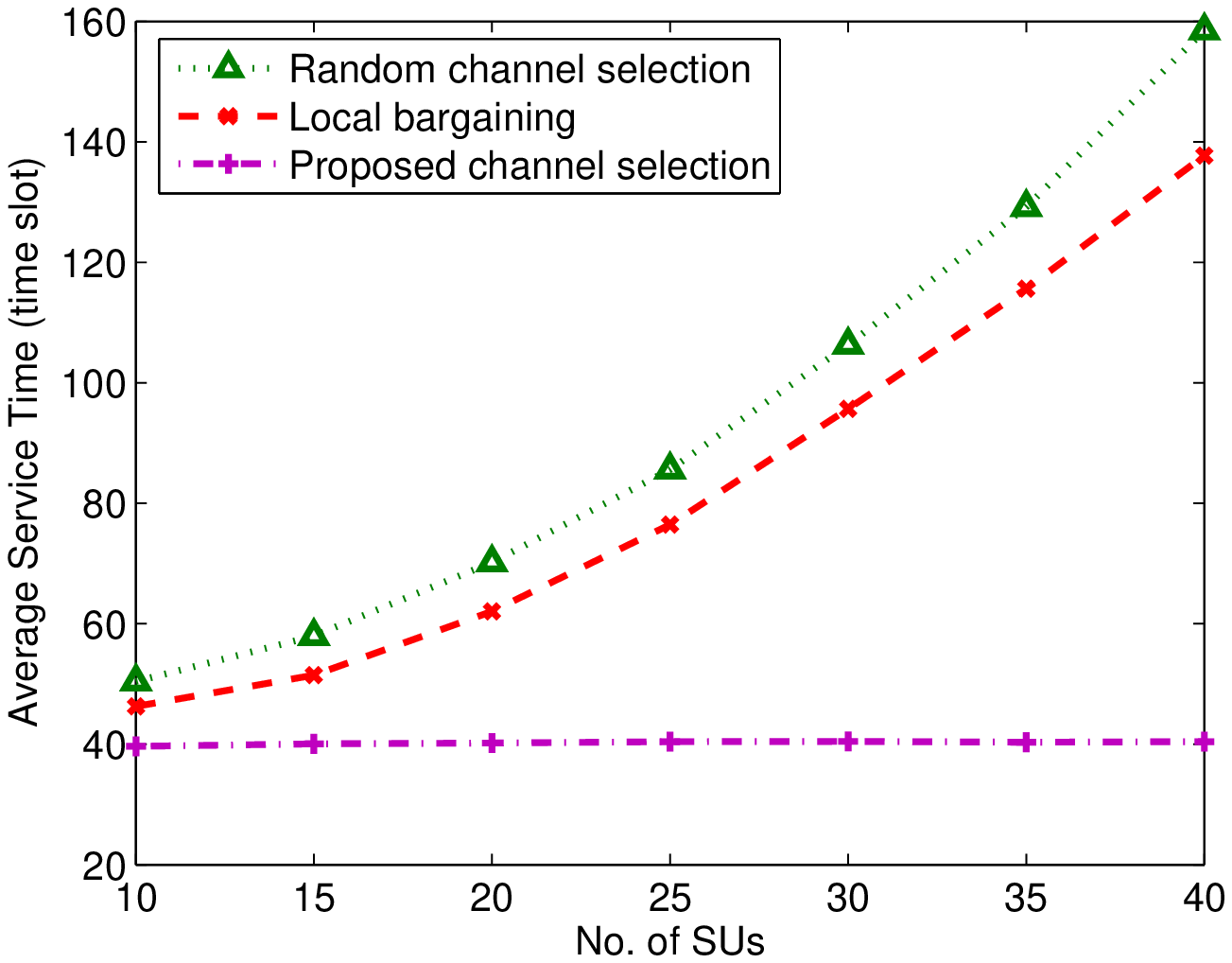}\label{fig:ast_nodes}}
\caption{\small Performance of the channel selection schemes in a multiple-pair-SU scenario under varying number of SUs.}
\label{fig:channel_selection_multisu}
\end{figure}

Fig. \ref{fig:s_nodes} and Fig. \ref{fig:ast_nodes} show the performance under different number of SUs, when there are 10 channels and the SU and PU traffic load is $500$ packet/second and $10$ packet/second, respectively. In Fig. \ref{fig:channel_selection_multisu}, we only show the local bargaining method, random channel selection, and the proposed channel selection. We exclude the greedy method because the greedy method constantly achieves zero throughput. Thus, its average service time is meaningless. As shown in the figures, the proposed channel selection scheme constantly achieves the highest throughput. This is because that the random channel selection scheme cannot eliminate collisions among SUs during spectrum handoffs. Additionally, in the local bargaining method, all SUs involved need to broadcast signaling messages twice in order to obtain a collision-free channel assignment, which leads to longer spectrum handoff delay and lower throughput.

\begin{figure}[hbt]
\centering
\subfigure[Number of collisions among SUs per second.]
{\includegraphics[width=0.49\textwidth]{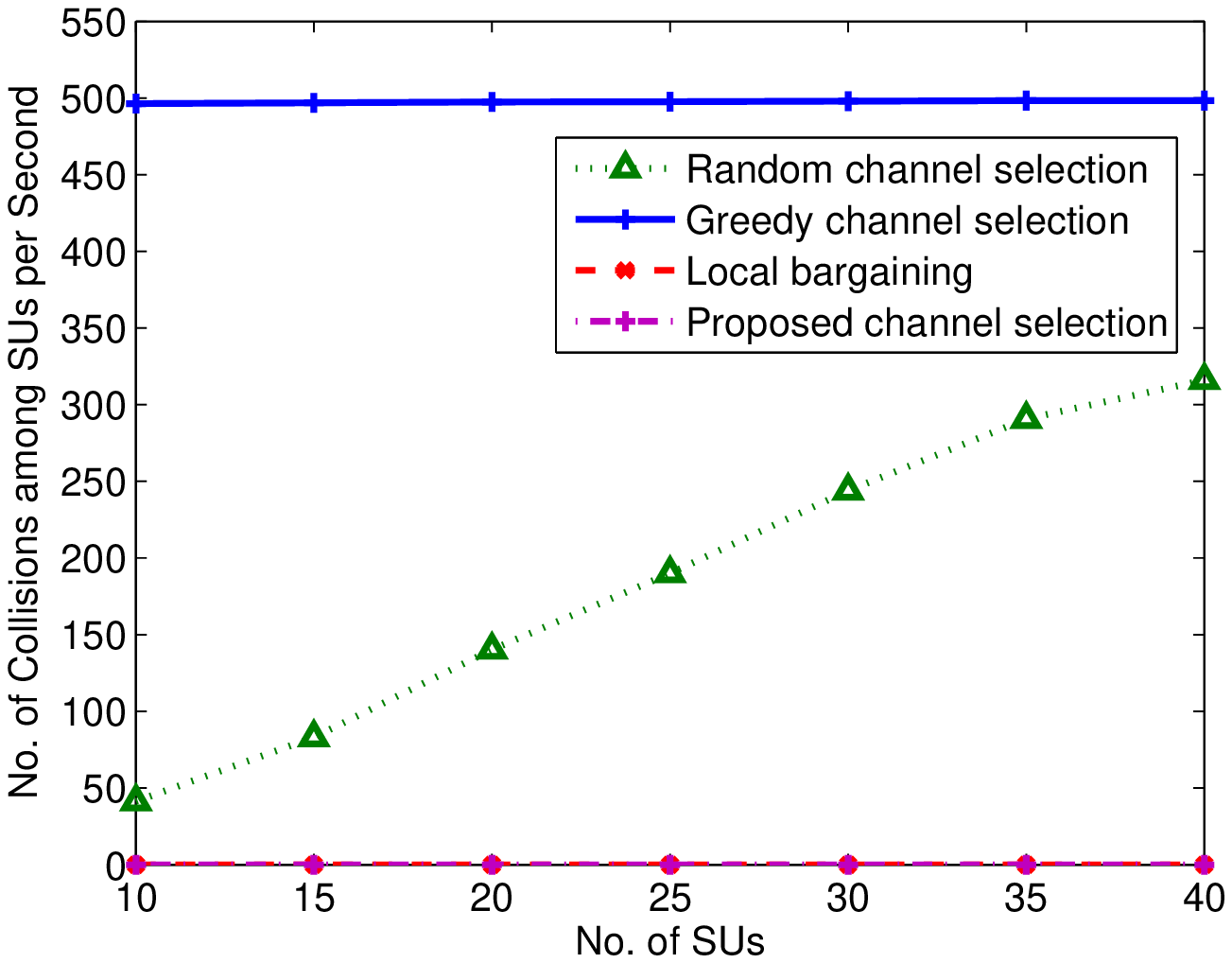}\label{fig:colsu_nodes}}
\subfigure[Average spectrum handoff delay.]
{\includegraphics[width=0.49\textwidth]{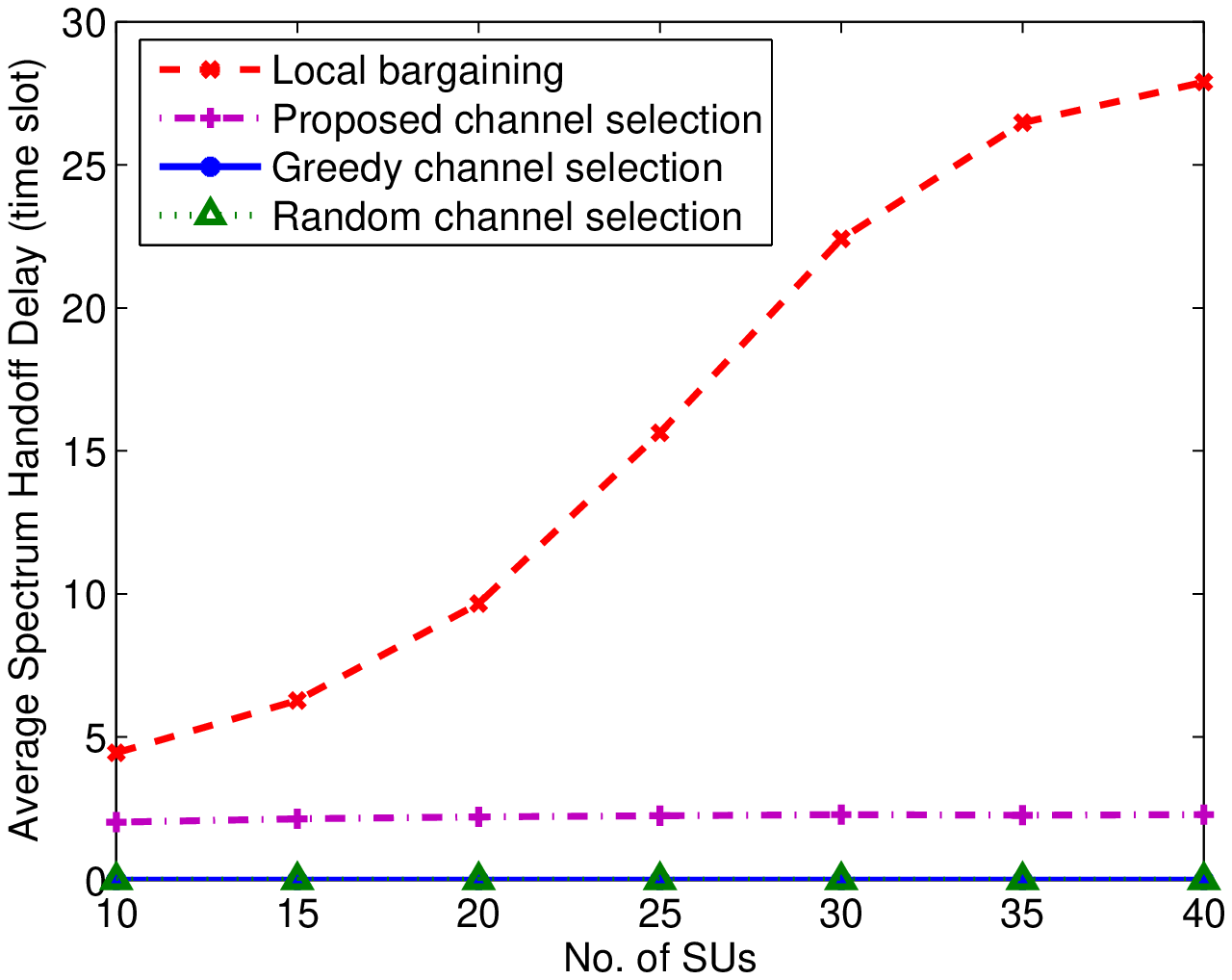}\label{fig:delay_nodes}}
\caption{\small Performance of the channel selection schemes in a multiple-pair-SU scenario under varying number of SUs.}
\label{fig:channel_selection_col_over}
\end{figure}

Fig. \ref{fig:channel_selection_col_over} shows the number of collisions among SUs per second and the average spectrum handoff delay of different channel selection schemes under varying number of SUs. It is shown in Fig. \ref{fig:colsu_nodes} that the greedy method and the random channel selection method cause more collisions among SUs than the local bargaining and the proposed channel selection method. While on the other hand, the local bargaining method cause much longer average spectrum handoff delay than the proposed channel selection scheme, as shown in Fig. \ref{fig:delay_nodes}. Therefore, the proposed channel selection scheme is the most suitable one for spectrum handoff scenarios.

\section{The Proposed Three Dimensional Discrete-time Markov Model}
\label{sc:analysis}
In this section, we develop a Markov model to analyze the performance of the reactive spectrum handoff process based on the single rendezvous coordination scheme. For simplicity, we assume that there are only two SUs in the network. We also ignore the propagation delay or any processing time in our analysis.
\subsection{The Proposed Markov Model}
\label{ssc:markov}
Based on the time slotted channels, any action of a SU can only be taken at the beginning of a time slot. In addition, the status of a SU in the current time slot only relies on its immediate past time slot. Such discrete-time characteristics allow us to model the status of a SU using Markov chain analysis. The status of a SU in a time slot can only be one of the following:
\begin{itemize}
	\item \textit{Idle}: no packet arrives at a SU.
	\item \textit{Transmitting}: the transmission of a SU does not collide with PU packets in a time slot, i.e., successful transmission.
	\item \textit{Collided}: the transmission of a SU collides with PU packets in a time slot, i.e., unsuccessful transmission.
	\item \textit{Backlogged}: a SU has a packet to transmit in the buffer but fails to access a channel.
\end{itemize}
Note that there are two cases that a SU can be in the \textit{Backlogged} status. In the first case, when a SU pair initiates a new transmission, if multiple SU pairs select the same channel for transmissions, a collision among SUs occurs and no SU pair can access the channel. Thus, the packet is backlogged. Similarly, in the second case, when a SU pair performs a spectrum handoff, if multiple SU pairs select the same channel, a collision among SUs occurs and the frame in each SU is also backlogged.

\begin{figure}[htb!]
	\centering
		\includegraphics[width=0.8\textwidth]{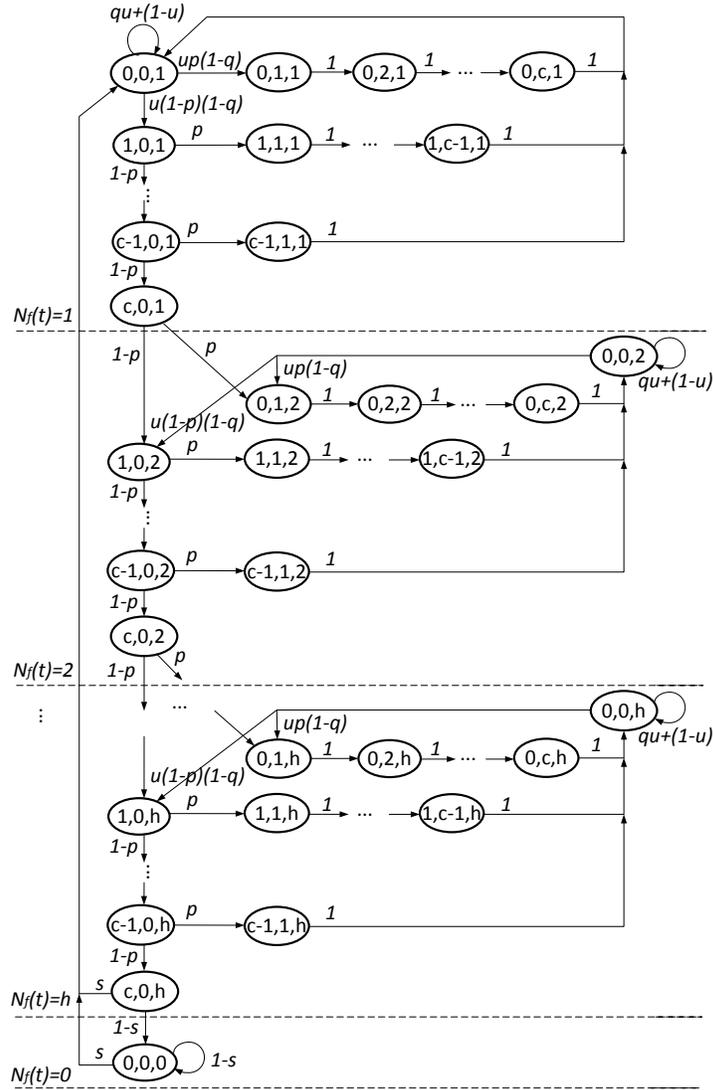}
	\caption{The transition diagram of the proposed Markov model.}
	\label{fig:markov}
\end{figure}

As mentioned in Section \ref{sc:introduction}, we consider the scenario that when a collision between a SU and PU happens, the overlapping of a SU frame and a PU packet is not negligible. Thus, the number of time slots that a SU frame collides with a PU packet is an important parameter to the performance of SUs. Based on the above analysis, the state of the proposed Markov model at time slot $t$ is defined by a vector $(N_t(t),N_c(t),N_f(t))$, where $N_t(t),N_c(t), {\rm and~} N_f(t)$ denote the number of time slots including the current slot that are successfully transmitted in the current frame, the number of time slots including the current slot that are collided with a PU packet in the current frame, and the number of frames that have been successfully transmitted plus the current frame that is in the middle of a transmission at time slot $t$, respectively. Therefore, $N_t(t)\!+\!N_c(t)\!\leq \!c$. Fig. \ref{fig:markov} shows the state transition diagram of our proposed three dimensional Markov chain. There are totally $(h\!+\!1)$ tiers in the state transition diagram. For each tier, it is a two dimensional Markov chain with a fixed $N_f(t)$. Table \ref{tb:notation} summarizes the notations used in our Markov model.
\begin{table}[htb]\caption{Notations Used in the Markov Analysis}
\centering
\begin{tabular}{cl}
\hline\noalign{\smallskip}
Symbol & Definition \\
\noalign{\smallskip}\svhline\noalign{\smallskip}
$p$ & Probability that a PU packet arrives in a time slot \\
$s$ & Probability that a SU packet arrives in a time slot \\
$h$ & Number of frames in a SU packet \\
$c$ & Number of time slots in a frame\\
$q$ & Probability of a collision among SUs\\
$u$ & Probability that at least one channel is idle \\
\noalign{\smallskip}\hline\noalign{\smallskip}
\end{tabular}
\label{tb:notation}
\end{table}

From Fig. \ref{fig:markov}, it is observed that the proposed Markov model accurately capture the status of a SU in a time slot. The state $(N_t(t)\!\!=\!\!0,N_c(t)\!\!=\!\!0,N_f(t)\!\!=\!\!0)$ in Fig. \ref{fig:markov} represents that a SU is in the $Idle$ status. Similarly, the states $(N_t(t)\!\in\!\![1,c],N_c(t)\!=\!0,N_f(t)\!\in\![1,h])$ represent the $Transmitting$ status, i.e., no collision. The states $(N_t(t)\in[0,c-1],N_c(t)\in[1,c],N_f(t)\in[1,h])$ represent the $Collided$ status. At last, the states $(N_t(t)\!\!=\!\!0,N_c(t)\!\!=\!\!0,N_f(t)\in[1,h])$ represent the $Backlogged$ status, where $(N_t(t)\!\!=\!\!0,N_c(t)\!\!=\!\!0,N_f(t)\!\!=\!\!1)$ is the \textit{Backlogged} status during a new transmission. As shown in Fig. \ref{fig:markov}, the feature of the common frequency-hopping sequence scheme is captured in our model that a SU can only start a new transmission when there is a channel available.\footnote{In the following discussion, we use the terms ``states'' in our proposed Markov model and the ``status'' of a SU in a time slot interchangeably. We also use the notations $(N_t(t\!+\!1)\!\!=\!\!i,N_c(t\!+\!1)\!\!=\!\!j,N_f(t\!+\!1)\!\!=\!\!k)$ and $(i,j,k)$ to represent a state interchangeably.}

\subsection{Derivation of Steady-State Probabilities}
\label{ssc:prob}
To obtain the steady-state probabilities of the states in the three dimensional Markov chain shown in Fig. \ref{fig:markov}, we first get the one-step state transition probability.\footnote{We denote the one-step state transition probability from time slot $t$ to time slot $t+1$ as $P(i_1,j_1,k_1|i_0,j_0,k_0)\!\!=\!\!P(N_t(t\!+\!1)\!\!=\!\!i_1,N_c(t\!+\!1)\!\!=\!\!j_1,N_f(t\!+\!1)\!\!=\!\!k_1|N_t(t)\!\!=\!\!i_0,N_c(t)\!\!=\!\!j_0,N_f(t)\!\!=\!\!k_0)$.} Thus, the non-zero one-step state transition probabilities for any $0\!<\!i_0\!<\!c, 0\!<\!j_0\!<\!c,{\rm and~} 0\!<\!k_0\!<\!h$ are given as follows:
\begin{equation}
\left\{
\begin{array}{ll}
P(0,0,k_0|0,0,k_0)=qu+(1-u) \\
P(1,0,k_0|0,0,k_0)=u(1-p)(1-q) \\
P(0,1,k_0|0,0,k_0)=up(1-q) \\
P(i_0,j_0+1,k_0|i_0,j_0,k_0)=1 \\
P(i_0,1,k_0|i_0,0,k_0)=p \\
P(i_0+1,0,k_0|i_0,0,k_0)=1-p\\
P(1,0,k_0+1|c,0,k_0)=1-p\\
P(0,1,k_0+1|c,0,k_0)=p\\
P(0,0,0|c,0,h)=1-s\\
P(0,0,1|c,0,h)=s\\
P(0,0,0|0,0,0)=1-s\\
P(0,0,1|0,0,0)=s
\end{array}
\right.
\end{equation}

Let $P_{(i,j,k)}\!\!=\!\!\lim_{t\to\infty}P(N_t(t)\!\!=\!\!i,N_c(t)\!\!=\!\!j,N_f(t)\!\!=\!\!k),i\!\in\![0,c],j\!\in\![0,c],k\!\in\![0,h]$ be the steady-state probability of the Markov chain. We first study a simple case where no PU exists in the CR network. Then, we consider the scenario where SUs coexist with PUs.

\subsubsection{Case One: No PU Exists in a Network}
\label{sssc:nopu}
In this case, since the probability that a PU packet arrives in a time slot is equal to zero (i.e., $p\!=\!0$), all channels are always available for SUs (i.e., $u$=1) and a SU does not need to perform spectrum handoffs during a data transmission. Thus, a SU cannot be in the \textit{Collided} state. In addition, a SU can only be in the \textit{Backlogged} state when it initiates a new transmission (i.e., the \textit{Backlogged} states are reduced to $(N_t(t)\!\!=\!\!0,N_c(t)\!\!=\!\!0,N_f(t)\!\!=\!\!1)$. Thus, the steady-state probabilities of the \textit{Transmitting} and \textit{Idle} state can be represented in terms of the steady-state probability of the \textit{Backlogged} state $P_{(0,0,1)}$. Hence, from Fig. \ref{fig:markov},
\begin{equation}\label{eq:eq1}
P_{(i,0,k)}=(1-q)P_{(0,0,1)}, {\rm ~for~} 1\leq i\leq c, 1\leq k\leq h,
\end{equation}
\begin{equation}\label{eq:eq2}
P_{(0,0,0)}=\frac{(1-s)(1-q)}{s}P_{(0,0,1)}.
\end{equation}
Since $\sum_i\sum_j\sum_kP_{(i,j,k)}\!\!=\!\!1$, we can calculate the steady-state probability of every state in the Markov chain. Note that the probability of a collision among SUs, $q$, depends on the channel selection scheme. The derivation of $q$ is given in Section \ref{sc:selection}.

\subsubsection{Case Two: SUs Coexist with PUs in a Network}
\label{sssc:pu}
If the probability that a PU packet arrives in a time slot is not equal to zero (i.e., $p\!\neq\!0$), collisions between SUs and PUs may occur when a SU transmits a frame. Thus, the steady-state probabilities of the \textit{Collided} states are not zero. Similar to the no-PU case, we represent the steady-state probabilities in terms of $P_{(0,0,1)}$. First of all, for the first tier in Fig. \ref{fig:markov}, we can obtain the steady-state probabilities of all the \textit{Transmitting} states in terms of $P_{(0,0,1)}$, that is,
\begin{equation}\label{eq:tier1trans}
P_{(i,0,1)}=u(1-q)(1-p)^{i}P_{(0,0,1)}, {\rm ~for~} 1\leq i\leq c.
\end{equation}
Then, for the \textit{Collided} states with $i=0$,
\begin{equation}\label{eq:tier1col1}
P_{(0,j,1)}=up(1-q)P_{(0,0,1)}, {\rm ~for~} 1\leq j\leq c.
\end{equation}
For the \textit{Collided} states with $i>0$,
\begin{equation}\label{eq:tier1col2}
P_{(i,j,1)}\!=\!u(1\!-\!q)p(1\!-\!p)^{i}P_{(0,0,1)}, {\rm ~for~} 1\!\leq\! i\!\leq\! c\!-\!1, 1\!\leq\! j\!\leq\! c.
\end{equation}
For the $k$-th $(k>1)$ tier, we first derive $P_{(1,0,k)}$ and $P_{(0,1,k)}$:
\begin{equation}\label{eq:tierktran1}
P_{(1,0,k)}=(1-p)P_{(c,0,k-1)}+u(1-p)(1-q)P_{(0,0,k)},
\end{equation}
\begin{equation}\label{eq:tierkcol1}
P_{(0,1,k)}=pP_{(c,0,k-1)}+up(1-q)P_{(0,0,k)}.
\end{equation}
Then, the steady-state probabilities of the \textit{Transmitting} states when $i>1$ can be represented as
\begin{equation}\label{eq:tierktrans}
P_{(i,0,k)}=(1-p)^{i-1}P_{(1,0,k)}, {\rm ~for~} 1<i\leq c.
\end{equation}
Similar to the derivation method for the first tier, for the \textit{Collided} states with $i=0$,
\begin{equation}\label{eq:tierkcol2}
P_{(0,j,k)}=P_{(0,1,k)}, {\rm ~for~} 1\leq j\leq c.
\end{equation}
For the \textit{Collided} states with $i>0$,
\begin{equation}\label{eq:tierkcol3}
P_{(i,j,k)}\!=\!p(1\!-\!p)^{i\!-\!1}P_{(1,0,k)}, {\rm ~for~} 1\!\leq\! i\!\leq\! c-1, 1\!\leq\! j\!\leq\! c.
\end{equation}
Then, for the $Backlogged$ state in the $k$-th tier,
\begin{equation}\label{eq:tierkback}
\sum_{i=0}^{c-1}P_{(i,c-i,k)}=u(1-q)P_{(0,0,k)}.
\vspace{-0.05in}
\end{equation}
Combining (\ref{eq:tierktran1}) through (\ref{eq:tierkback}), we obtain the following equations using basic mathematical manipulations:
\begin{equation}\label{eq:tierktrans1}
P_{(1,0,k)}=\frac{1}{(1-p)^{c-1}}P_{(c,0,k-1)},
\end{equation}
\begin{equation}\label{eq:tierkcol11}
P_{(0,1,k)}=\frac{p}{(1-p)^c}P_{(c,0,k-1)},
\end{equation}
\begin{equation}\label{eq:tierkback1}
P_{(0,0,k)}=\frac{1-(1-p)^c}{u(1-q)(1-p)^c}P_{(c,0,k-1)}.
\end{equation}
Then, from (\ref{eq:tierktrans}),
\begin{equation}\label{eq:recur1}
P_{(c,0,k-1)}=(1-p)^{c-1}P_{(1,0,k-1)}.
\vspace{-0.05in}
\end{equation}
Combining (\ref{eq:tierktrans1}) and (\ref{eq:recur1}), we find the following relationship:
\begin{equation}\label{eq:recur2}
P_{(c,0,k)}=P_{(c,0,k-1)}.
\end{equation}
Thus,
\begin{equation}
\vspace{-0.05in}
\label{eq:recur3}
P_{(c,0,k)}=u(1-q)(1-p)^cP_{(0,0,1)}.
\end{equation}
(\ref{eq:recur3}) indicates the steady-state probabilities of the states in the $k$-th tier are independent of $k$. Now, we have all the steady-state probabilities of the states in all tiers except the state $(0,0,0)$. At last, for the $Idle$ state,
\begin{equation}\label{eq:idle}
P_{(0,0,0)}=\frac{1-s}{s}u(1-q)(1-p)^cP_{(0,0,1)}.
\end{equation}
Similarly, since $\sum_i\sum_j\sum_kP_{(i,j,k)}=1$, we can get the steady-state probability of every state in the Markov chain. If we denote $\Theta$ as the normalized throughput of SU transmissions, $\Theta$ is the summation of the steady-state probabilities of all the $Transmitting$ states in our proposed Markov model. That is,
\begin{equation}\label{eq:throu}
\vspace{-0.05in}
\Theta=\sum_{k=1}^h\sum_{i=1}^cP_{(i,0,k)}. 
\end{equation}

\subsection{The Probability that at Least One Channel is Idle}
\label{ssc:puchannel}
In the above derivations, $u$ and $q$ are unknown. In this subsection, we calculate the probability that at least one channel is idle, $u$. Without loss of generality, we associate a PU with one channel and model the activity of a PU on a channel as an ON/OFF process \cite{Zhang06}\cite{Su08}. SUs can only exploit the channels when the channels are idle (i.e., in the OFF period). We assume that the buffer in each PU can store at most one packet at a time. Once a packet is stored at a buffer, it remains there until it is successfully transmitted. Thus, we assume that the OFF period of a channel follows the geometric distribution, where the probability mass function (pmf) is given by 
\begin{equation}
\Pr(N_{OFF}=n)=p(1-p)^{n},
\label{eq:interarrival}
\end{equation}
where $N_{OFF}$ is the number of time slots of an OFF period.

\begin{figure}[htb!]
	\centering
		\includegraphics[width=0.6\textwidth]{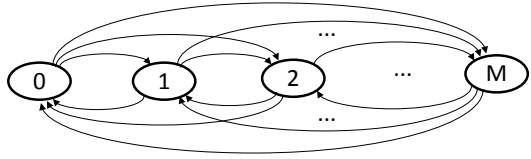}
	\caption{The transition diagram of the number of channels used by PUs in one time slot.}
	\label{fig:markov2}
\end{figure}

Let $\Omega(t)$ be the number of channels used by PUs at time slot $t$. The process $\{\Omega(t),t=0,1,2,\cdots\}$ forms a Markov chain whose state transition diagram is given in Fig. \ref{fig:markov2}, in which the self loops are omitted. To characterize the behavior of the PU channels, we define $\mathcal{D}_\alpha^l$ as the event that $l$ PUs finish their transmissions given that there are $\alpha$ PUs in the network in a time slot. We also define $\mathcal{A}_\gamma^m$ as the event that $m$ PUs start new transmissions given that there are $\gamma$ idle PUs in a time slot. Thus, the probabilities of events $\mathcal{D}_\alpha^l$ and $\mathcal{A}_\gamma^m$ are:
\begin{equation}\label{eq:eventa}
\Pr(\mathcal{D}_\alpha^l)=\binom{\alpha}{l}v^l(1-v)^{\alpha-l},
\end{equation}
\begin{equation}\label{eq:eventb}
\Pr(\mathcal{A}_\gamma^m)=\binom{\gamma}{m}p^m(1-p)^{\gamma-m},
\end{equation}
where $v$ is the probability that a PU finishes its transmission in a slot. If the average length of a PU packet is denoted as $\bar{L}$, then $v\!=\!1/\bar{L}$. Therefore, the state transition probability from state $\{\Omega(t)\!=\!a\}$ to state $\{\Omega(t\!+\!1)\!=\!b\}$ can be written as
\begin{equation}
p_{ab} = 
\left\{
\begin{array}{ll}
\sum_{l=0}^a\Pr(\mathcal{D}_a^l)\Pr(\mathcal{A}_{M-a+l}^{b-a+l}), &{\rm for~} b\geq a \\
\sum_{l=a-b}^a\Pr(\mathcal{D}_a^l)\Pr(\mathcal{A}_{M-a+l}^{b-a+l}), &{\rm for~} b< a.
\end{array}
\right.
\label{eq:trans}
\end{equation}
Therefore, we can obtain the steady-state probabilities of the number of busy channels in the band in a time slot, denoted as $\textbf{g}=[g_0 ~~ g_1~~ g_2~\cdots~ g_M]^T$, where $g_i$ denotes the steady-state probability that there are $i$ busy channels in a time slot. Hence, $u=\sum_{i=0}^{M-1}g_i$.

\subsection{Results Validation}
\label{ssc:validation}
\begin{figure}[htb!]
	\centering
		\includegraphics[width=0.65\textwidth]{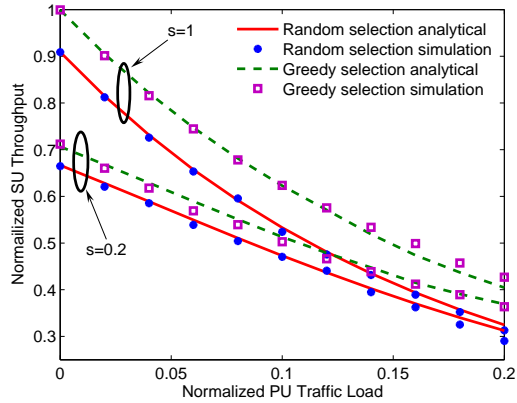}
	\caption{Analytical and simulation results of the normalized SU throughput in a two-SU scenario.}
	\label{fig:twousers}
\end{figure}

In this subsection, we validate the numerical results obtained from our proposed Markov model using simulation. Note that we only consider two SUs in the network, the probability of collision among SUs is always zero (i.e., $q\!=\!0$). Thus, we validate our numerical results in a two-SU scenario, where the number of PU channels, $M\!=\!10$. The number of frames in a SU packet, $h\!=\!1$, and the number of slots in a frame, $c\!=\!10$. We assume that the SU packets are of fixed length. Thus, $\sigma\!=\!1/(ch)$. Fig. \ref{fig:twousers} depicts the analytical and simulation results of the normalized SU throughput using the random channel selection scheme and the greedy channel selection scheme. It can be seen that the simulation results match extremely well with the numerical results in both schemes with the maximum difference only $3.84\%$ for the random selection and $4.09\%$ for the greedy selection. It is also shown that, under the same SU traffic load, the greedy channel selection scheme always outperforms the random channel selection scheme in terms of higher SU throughput.

\subsection{The Impact of Spectrum Sensing Delay}
\label{sc:sensing}
In this section, we investigate the impact of the spectrum sensing delay on the performance of a spectrum handoff process. The spectrum sensing delay considered in this chapter is defined as the duration from the moment that a collision between a SU and PU happens to the moment that the SU detects the collision (i.e., the overlapping time between a SU and PU transmission). Let $T_s$ be the spectrum sensing delay. Therefore, a SU does not need to wait till the last time slot of a frame to realize the collision. It only needs to wait for $T_s$ to realize that a collision with a PU packet occurs and stops the current transmission immediately. In a recent work \cite{CWWangGC10}, the spectrum sensing time is considered as a part of the spectrum handoff delay. However, the definition of the spectrum sensing time in \cite{CWWangGC10} is different from the definition considered in this chapter. In \cite{CWWangGC10}, the spectrum sensing time only refers to the duration that a SU finds an available channel for transmission after a collision occurs. Thus, the spectrum sensing time can be as low as zero in \cite{CWWangGC10}. In addition, the overlapping time of a SU and PU collision is neglected in \cite{CWWangGC10}. However, the spectrum sensing delay considered in this chapter is not negligible. 

\begin{figure}[htb!]
	\centering
		\includegraphics[width=0.65\textwidth]{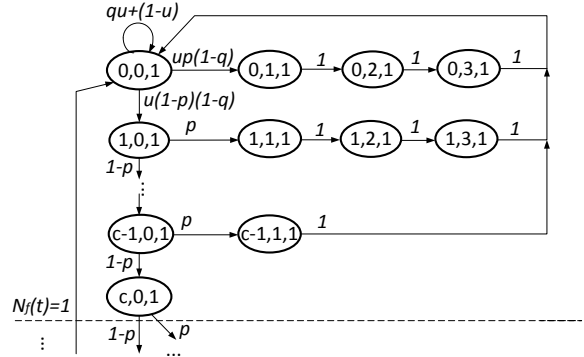}
	\caption{The modified Markov model based on the spectrum sensing delay when $T_s$ equals 3 time slots.}
	\label{fig:sensingdelay}
\end{figure}

The spectrum sensing delay, $T_s$, can be easily implemented in our proposed three dimensional Markov model with minor modifications. Fig. \ref{fig:sensingdelay} shows the first tier of the modified three dimensional discrete-time Markov chain when $T_s$ equals 3 time slots. It is shown that, for a fixed $N_t(t)$, the maximum number of $Collided$ states is $T_s$. The modified model of other tiers is similar to the first tier as shown in Fig. \ref{fig:sensingdelay}.

Compared with the original Markov model shown in Fig. \ref{fig:markov}, the derivation of the steady-state probabilities of the Markov model implemented with the spectrum sensing delay is exactly the same. The only difference is that the total number of the \textit{Collided} states in the modified Markov model is reduced from $[c(c\!+\!1)/2]h$ in the original Markov model to $[T_s(c\!-\!T_s\!+\!1)\!+\!T_s(T_s\!-\!1)/2]h$.

\begin{figure}[htb!]
	\centering
		\includegraphics[width=0.65\textwidth]{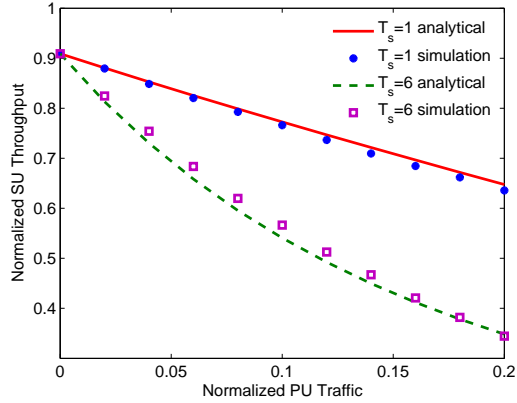}
	\caption{Analytical and simulation results of the normalized SU throughput under different spectrum sensing delay.}
	\label{fig:sensing}
\end{figure}

Fig. \ref{fig:sensing} shows the impact of the spectrum sensing delay on the SU throughput performance. We consider a two-SU scenario with different spectrum sensing delay using the random channel selection scheme. It is shown that the numerical results and analytical results match well with the maximum difference $1.83\%$ for $T_s\!=\!1$ and $4.56\%$ for $T_s\!=\!6$. It reveals that our proposed model can accurately predict the SU throughput. 

\section{Conclusion}
\label{sc:conclusion}
In this chapter, a proactive spectrum handoff framework in a CR ad hoc network scenario without the existence of a CCC is proposed. Compared with the sensing-based reactive spectrum handoff approach, the proposed framework can achieve fewer disruptions to primary transmissions by letting SUs proactively predict the future spectrum availability and perform spectrum handoffs before a PU occupies the current spectrum. We incorporated a single rendezvous and a multiple rendezvous network coordination scheme into the spectrum handoff protocol design, thus our proposed spectrum handoff framework is suitable for the network scenarios that do not need a CCC. Furthermore, most of the prior work on channel selection in spectrum handoffs only considers a two-SU scenario, while the channel selection issue for a multi-SU scenario is ignored. We also proposed a novel fully distributed channel selection scheme which leads to zero collision among SUs in a multi-SU scenario. Simulation results show that our proposed channel selection scheme outperforms the existing methods in terms of higher throughput and shorter handoff delay in multi-SU scenarios. 

Furthermore, a novel three dimensional discrete-time Markov chain is proposed to analyze the performance of SUs in the reactive spectrum handoff scenario in a two-SU CR ad hoc network is proposed. We performed extensive simulations in different network scenarios to validate our proposed model. The analysis shows that our proposed Markov model is very flexible and can be applied to various practical network scenarios.

\end{document}